\documentclass{article}
\usepackage[margin=1in]{geometry}
\usepackage{amsmath,amssymb}
\usepackage{float}
\usepackage{caption}
\usepackage[T1]{fontenc}
\usepackage[utf8]{inputenc}
\usepackage{textcomp}
\usepackage{lmodern}
\usepackage{xcolor}
\usepackage{graphicx}
\PassOptionsToPackage{hyphens}{url}
\usepackage[unicode]{hyperref}
\IfFileExists{xurl.sty}{\usepackage{xurl}}{}
\IfFileExists{microtype.sty}{\usepackage{microtype}}{}
\IfFileExists{parskip.sty}{\usepackage{parskip}}{}

\makeatletter
\def\maxwidth{\ifdim\Gin@nat@width>\linewidth\linewidth\else\Gin@nat@width\fi}
\def\maxheight{\ifdim\Gin@nat@height>\textheight\textheight\else\Gin@nat@height\fi}
\makeatother
\setkeys{Gin}{width=\maxwidth,height=\maxheight,keepaspectratio}
\makeatletter
\def\fps@figure{htbp}
\makeatother
\providecommand{\tightlist}{%
  \setlength{\itemsep}{0pt}\setlength{\parskip}{0pt}}
\hypersetup{
  pdftitle={Forgeable Confirmation in Automated Computer Security Testing: Deterministic Rules versus AI Judges},
  hidelinks,
  pdfcreator={pdfLaTeX}}

\title{Forgeable Confirmation in Automated Computer Security Testing:
Deterministic Rules versus AI Judges}
\author{
\begin{tabular}{c}
Akihisa Fujiyama\textsuperscript{1,2}, Niwase Shamim\textsuperscript{1,*}\\[0.5em]
\small \textsuperscript{1}OmiCore Inc., Fukuoka, Japan\\
\small \textsuperscript{2}Graduate School of Engineering, Kyushu University, Fukuoka, Japan\\
\small \textsuperscript{*}Corresponding author: Niwase Shamim (contact@omi-core.com)
\end{tabular}
}
\date{}

\begin{document}
\maketitle

\begin{abstract}
AI is increasingly used to automate computer security testing, and the
tools must decide for themselves whether an attack succeeded. A finding
that a deterministic rule \emph{confirms} by observation is reported as
fact, whereas one that an LLM \emph{judges} exploitable is treated as an
opinion. We ask whether the system under test can forge that
confirmation. In offline security testing of a four-stage AI-assisted
pipeline, nine of its fifteen confirmation mechanisms are forgeable, and
forgeability is predicted entirely by whether the decision reads
attacker-controlled data. We formalise this as an \emph{auditable attack
surface} and test it prospectively: on sixteen held-out mechanisms,
predictions fixed before any attack separated forgeable from unforgeable
mechanisms exactly (Fisher \(p = 5.5 \times 10^{- 4}\); 13 of 16 under
the originally specified adversary), and across 12,203 mechanisms in
public scanner templates the prediction was 99.9~\% accurate.
Deterministic rules proved \emph{cheaper} to forge than eight
open-weight LLM judges, failing at 2~\% of attacker-controlled response
content against a median of 50~\%. No implementation of one check was
both robust and precise, and routing between a rule and an AI judge
raised forgery to 99~\%. Moving the decisive evidence to a channel the
attacker cannot write cuts attack success from 97~\% to 0~\%, and an
\emph{escalate} verdict recovers the sensitivity this costs. The
protection fails when the scanned host is itself the adversary. The
results bear on AI security agents and on benchmarks that score success
by string matching.
\end{abstract}


\section{Introduction}\label{introduction}

Security-operations analysts triage large volumes of machine-generated
findings; practitioners report that most are false and that manual
validation dominates their workload\textsuperscript{1}. The problem
predates security operations centres. Developers abandon static-analysis
tools with high false-positive rates\textsuperscript{2,3}, tool builders
treat the false-positive budget as the primary design
constraint\textsuperscript{4}, security-focused static analysis shows
the same pattern\textsuperscript{5}, and black-box web scanners have
long missed entire vulnerability classes while emitting findings that a
human must adjudicate\textsuperscript{6,7}.

Verifier-gated pipelines address this by attaching a different kind of
evidence to some findings. A finding may carry a model\textquotesingle s
judgement, or it may carry an \emph{observation}: for example, a forged
token returned the same protected resource, and the two response bodies
are hash-identical. Pipelines label such findings accordingly
(\emph{confirmed by observation} rather than \emph{84~\% likely}). When
queues are ordered, bug-bounty reports filed or deployment gates opened
on that label, an attacker who can manufacture it also manufactures the
priority.

The question is becoming more pressing. Language-model agents exploit
real one-day vulnerabilities\textsuperscript{8}, automate
penetration-testing workflows\textsuperscript{9}, and are evaluated on
capture-the-flag and real-world exploitation
benchmarks\textsuperscript{10,11}. Each of these systems decides whether
an exploit succeeded by reading bytes that the target produced.

We study a four-stage pipeline (generate~$\rightarrow$~judge~$\rightarrow$~confirm~$\rightarrow$~measure) in
which the confirmation stage is the verdict authority, and ask whether
its confirmations can be forged. Our contributions are: (i) the
\emph{auditable attack surface}, a criterion that predicts from source
code which confirmation mechanisms are forgeable, derived on one
pipeline, tested prospectively on held-out modules and applied to a
third-party corpus; (ii) measurements of attacker cost for deterministic
rules and LLM judges, both as a robustness curve over attacker budgets
and as an exact minimum payload length; (iii) evidence from six
independent implementations, and from rule--judge composition, that the
weakness lies in the task definition rather than in any one
implementation; and (iv) a defence, its measured cost, and the adversary
model under which it holds.

\section{Background}\label{background}

\textbf{Attack surface of a decision.} Attack-surface measurement for
whole systems is well established: Manadhata and Wing formalise
attackability along method, data and channel
dimensions\textsuperscript{12}, and a systematic review documents the
many definitions in use\textsuperscript{13}. We apply the idea to a
single decision procedure, so that two verifiers can be compared by what
an attacker must supply rather than by their stated confidence. To our
knowledge, the attack surface of the check that decides whether an
attack succeeded has not been measured. The difficulty of trusting a
tool to certify its own output is long recognised\textsuperscript{14};
in software supply chains the current response is to attach verifiable
provenance to each step\textsuperscript{15}.

\textbf{LLM-as-a-judge.} Using one model to score
another\textquotesingle s output is standard practice with documented
limitations: position, verbosity and self-enhancement
biases\textsuperscript{16}, sensitivity to answer
order\textsuperscript{17}, preference for a model\textquotesingle s own
generations\textsuperscript{18}, and a broader design
space\textsuperscript{19}. Two common mitigations are used in our
pipeline: repeated sampling of one judge\textsuperscript{20} and a panel
of smaller judges\textsuperscript{21}. Judges can be flipped by a single
token, with false-positive rates up to 80~\%\textsuperscript{22}, and
such flips can be found by optimisation\textsuperscript{23}. We transfer
this attack to a security setting: the transfer to our judges is weak
(at most 20~\% of cases flipped, and three of five judges never flip),
whereas the deterministic marker rule of the same pipeline flips on
100~\% (Fig.~S1d).

\textbf{Prompt injection.} Indirect prompt injection compromises
LLM-integrated applications through retrieved
content\textsuperscript{24,25}; attacks and defences have been
formalised and benchmarked\textsuperscript{26}, including for tool-using
agents\textsuperscript{27,28}. These attacks exploit a system that does
not separate data from instructions. A rule that searches a response
body for marker strings has the analogous flaw without any model: it
does not separate data from \emph{verdict}. Consistent with this, the
prompt-injection payload family wins no case against any deterministic
arm in our search, while a three-character literal wins most.

\textbf{Adversarial machine learning.} Evasion of learned security
classifiers is well studied\textsuperscript{29}. Two methodological
lessons apply here: defences must be evaluated against adaptive
attackers\textsuperscript{30,31}, and attacks must be realisable in the
problem space\textsuperscript{32}. Evaluation bias is a known hazard in
security machine learning\textsuperscript{33,34}. Our budget sweep
applies the same discipline to rules: every payload is a byte sequence
that a server could return, and a deletion-only control separates
forgery from loss of evidence.

\textbf{Ground truth.} Fuzzing evaluations are sensitive to
methodological choices\textsuperscript{35}, which motivated ground-truth
benchmarks with injected bugs\textsuperscript{36}. Vulnerability
datasets carry 20--71~\% inaccurate labels\textsuperscript{37},
performance on curated data does not transfer to realistic
data\textsuperscript{38,39}, and the effect persists for code language
models\textsuperscript{40}; guidelines for sound experimental design are
long established\textsuperscript{41}. The problem applies to this study
directly, because our positive class is labelled by the mechanisms under
examination; we discuss this in the Limitations.

\textbf{Exploitation, reliance and measurement.} Automatic exploit
generation includes a verification step whose signal is not
attacker-authored: a segmentation fault is observed by the operating
system, not asserted by the target\textsuperscript{42,43}. LLM-driven
offensive tools instead verify success from application
responses\textsuperscript{8,9}, on benchmarks whose success criteria are
string matches\textsuperscript{10,11}. How operators use automated
verdicts is well studied: automation is misused and disused in
predictable ways\textsuperscript{44}, explanations increase acceptance
regardless of correctness\textsuperscript{45}, cognitive forcing reduces
overreliance\textsuperscript{46}, reliance is measured with
appropriateness constructs\textsuperscript{47,48}, and hedged advice is
followed less than unhedged advice of equal
accuracy\textsuperscript{49}. The label \emph{confirmed by observation}
is therefore consequential. We measure the machine rather than the
operator (Fig.~S3). Agreement is reported
chance-corrected\textsuperscript{50--52}. The relevant failure mode is
specification gaming, in which a proxy diverges from its goal under
optimisation\textsuperscript{53--56}. Vulnerability prioritisation
systems already condition on exploitation
likelihood\textsuperscript{57,58}, so a forgeable \emph{confirmed} flag
propagates into them.

\section{Results}\label{results}

\subsection{Forgeability tracks what a decision reads, not whether it
is
deterministic}\label{forgeability-tracks-what-a-decision-reads-not-whether-it-is-deterministic}

\includegraphics[width=6.1in,height=5.83478in]{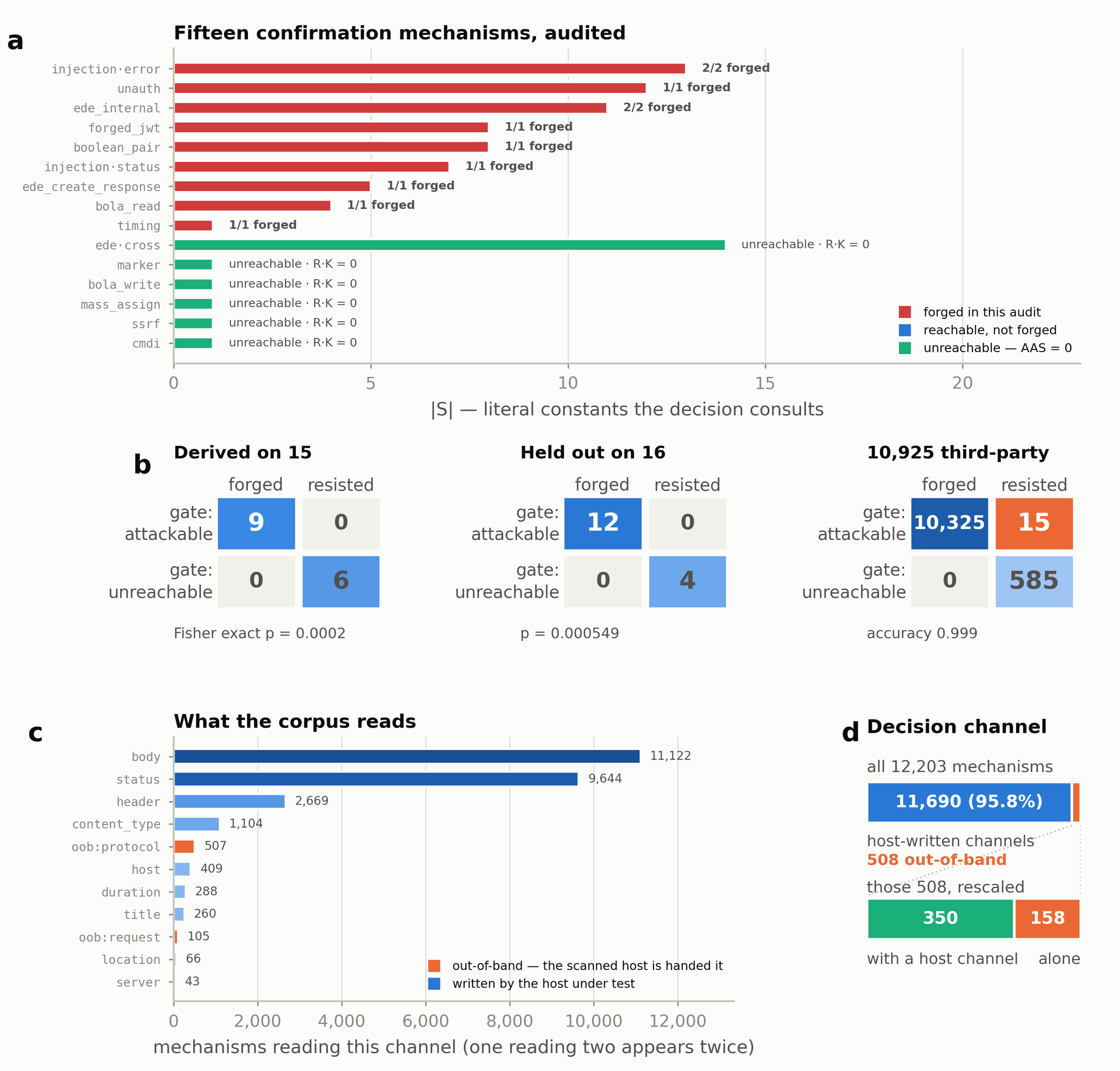}

\textbf{Figure 1. The reachability gate at three scales.} \textbf{(a)}
The fifteen deterministic confirmation mechanisms of the pipeline,
ordered by outcome and then by \(\mid S \mid\). Bar length is the number
of literal constants the decision consults; colour is the audit result
(red, forged; green, unreachable, \(G = 0\); blue, reachable but not
forged, which is empty here). Annotations give forged over attempted
attacks. Bar length and outcome are nearly unrelated: ede$\cdot$cross consults
fourteen constants and cannot be attacked, whereas timing consults one
and is forged. \textbf{(b)} The gate as a prediction. Rows are the
gate\textquotesingle s call and columns the attack outcome. \emph{Left}:
the 15 mechanisms on which the criterion was defined (Fisher exact
\(p = 2 \times 10^{- 4}\)). \emph{Centre}: 16 held-out mechanisms from
six modules, including an Active Directory scanner, predicted from
source and hash-pinned before any attack (\(p = 5.5 \times 10^{- 4}\)).
\emph{Right}: 10,925 third-party mechanisms, with forgeries constructed
and evaluated automatically (accuracy 0.999). The bottom-right cell
holds by definition; the informative cells are the top row, where 15 of
10,340 predicted-attackable mechanisms resisted, and the bottom-left
cell (predicted unreachable but forged), which is zero at all three
scales. \textbf{(c)} Decision channels of the 12,203 third-party
mechanisms; a mechanism reading two channels is counted twice. Orange
marks the two out-of-band channels, which the scanned host nonetheless
receives. \textbf{(d)} \emph{Top}: 95.8~\% of mechanisms read only
channels authored by the scanned host. \emph{Bottom}: of the 508 that
consult an out-of-band channel, 350 combine it with a host-written
channel, the structure of the hardened policy in eq.~(7), and 158 rely
on the out-of-band anchor alone.

We audited all fifteen deterministic confirmation mechanisms of the
pipeline (Fig.~1a, Table~1). Nine are forgeable and six are unreachable,
and the number of constants a mechanism consults does not determine the
outcome: a mechanism consulting fourteen constants cannot be attacked,
whereas one consulting a single threshold is forged (Fig.~1a). Every
forgeable mechanism decides by reading text or structure the attacker
wrote; every unreachable one relies on a tester-generated secret, an
out-of-band callback or a server-side read-back.

\textbf{Table 1. Deterministic confirmation mechanisms of the pipeline,
audited for forgeability.} R = 1 when attacker-controlled input reaches
the decision; K = 1 when the decisive value is attacker-knowable; AAS =
\textbar S\textbar{} when R$\cdot$K = 1 and 0 otherwise. Of the 15 mechanisms,
9 are forgeable and 6 are unreachable.

\includegraphics[width=6.1in,height=4.72154in]{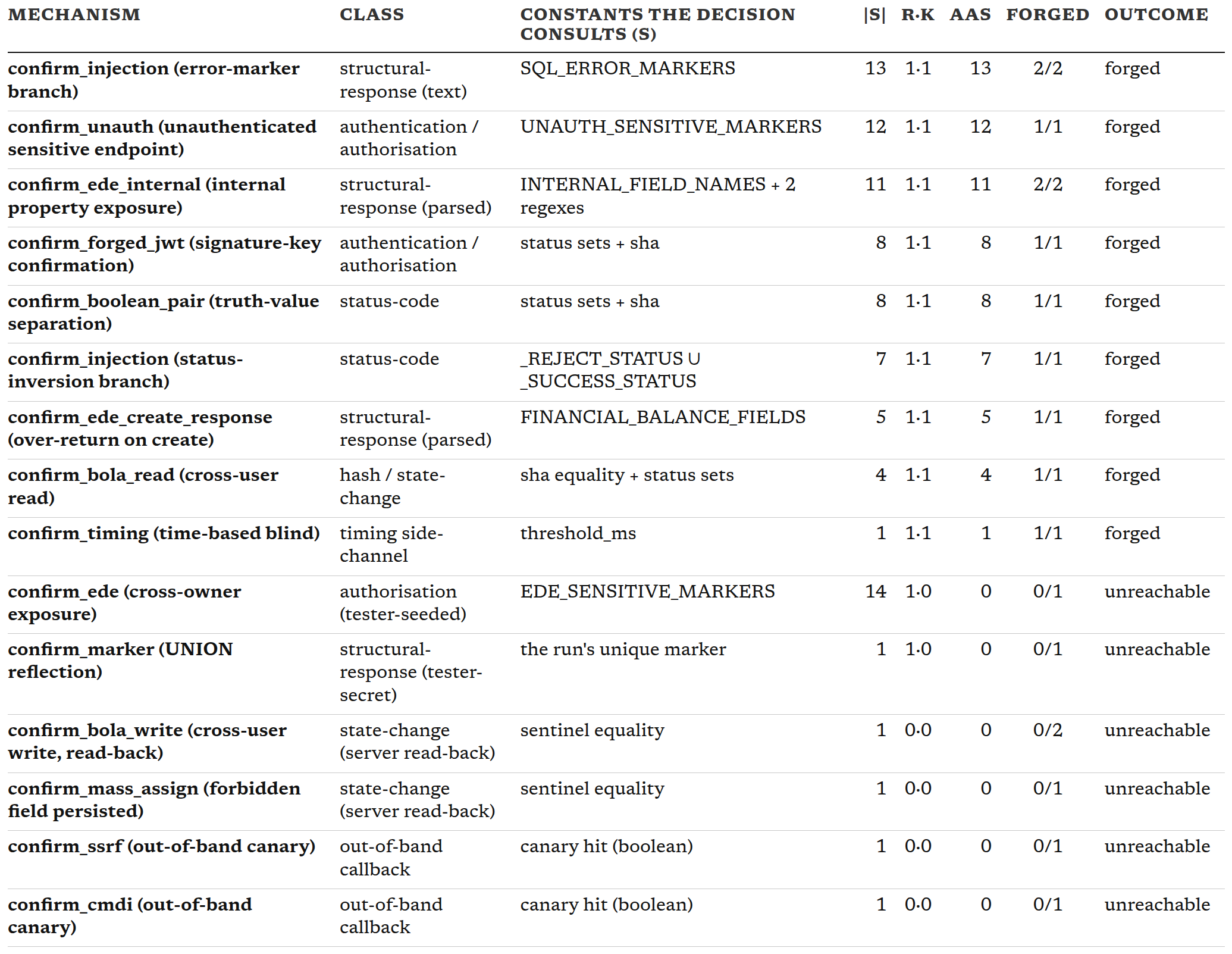}

Note. Overall attack success 61\%. Every mechanism passed its
clean-accuracy check before it was attacked.

\textbf{Held-out prediction.} A criterion that separates the cases it
was defined on may simply fit them, so we tested it prospectively
(Table~2): for sixteen mechanisms from six modules not used in its
derivation, predictions were read from source and hash-pinned before any
attack was written. All twelve mechanisms predicted forgeable were
forged and all four predicted unreachable resisted
(\(p = 5.5 \times 10^{- 4}\); Fig.~1b, centre), compared with
\(p = 2 \times 10^{- 4}\) on the fifteen mechanisms from which the
criterion was derived (Fig.~1b, left). The adversary model was widened
after one misprediction, to allow an injected line break, and the wider
model was then applied to all seven anchors of that module. Under the
original, narrower model the accuracy is 81.2~\% (\(p = 0.019\)); we
report both.

\textbf{Table 2. Held-out prediction test of the reachability gate.} 16
confirmation mechanisms from 6 modules not used to derive the criterion,
including an Active Directory scanner. Predictions were read from source
and hash-pinned (PREDICTION\_SHA = a41c7a1fec100b9f) before any attack
was written.

\includegraphics[width=6.1in,height=3.18991in]{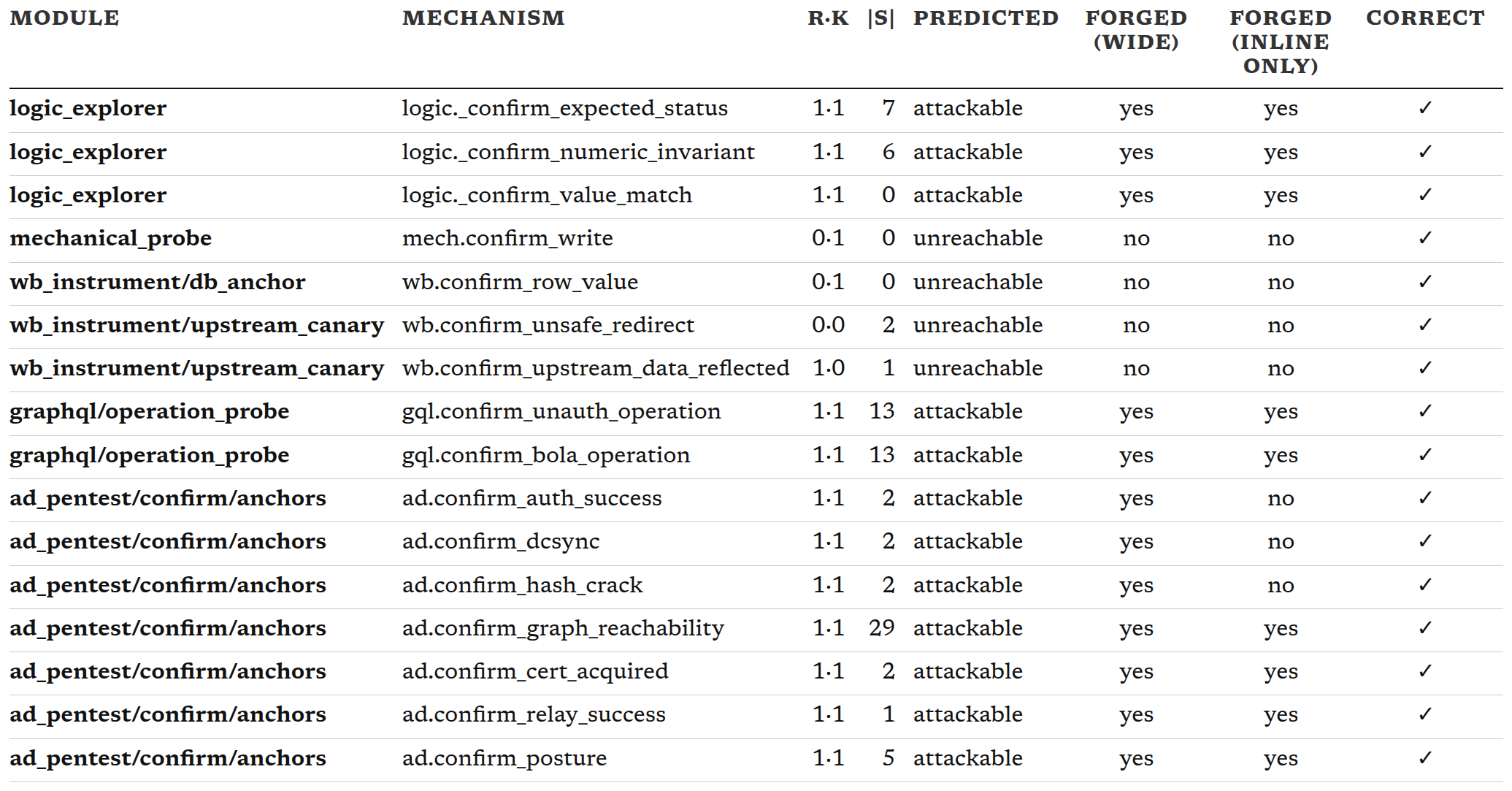}

Note. Wide adversary: 12 true positives, 0 false negatives, 0 false
positives, 4 true negatives --- accuracy 100\%, Fisher exact p =
0.000549. Narrower adversary (no injected line break): accuracy 81.2\%,
p = 0.019231. The wider adversary model was adopted after one
misprediction and then applied to all seven AD anchors; both scorings
are therefore reported. Deletion alone flips 0 of 16.

\textbf{Third-party corpus.} The six re-implementations reported below
address the concern that the result is specific to one codebase, but not
that it is specific to one set of authors. We therefore applied the gate
to 11,137 public scanner templates\textsuperscript{59} containing 12,203
confirmation mechanisms written by 1,611 contributors, scoring each
automatically from its matcher specification (Fig.~1b right, 1c, 1d;
Table~S7). Before any attack, 121 mechanisms were excluded because they
already fired on a benign placeholder (the clean-accuracy control of
eq.~(3)); 1,102 offered no writable matcher and 55 could not be
evaluated from their specification. Of the remaining 10,925, 10,325 were
forged, 15 resisted and 585 were out of the attacker\textquotesingle s
reach; no mechanism predicted unreachable was forged (accuracy 0.999;
Fig.~1b, right). Response body and status code dominate the channels
these mechanisms read (Fig.~1c), and 95.8~\% of mechanisms read only
channels authored by the scanned host; of the 508 that consult an
out-of-band channel, 350 combine it with a host-written channel, the
structure of the hardened policy of eq.~(7), and 158 rely on the
out-of-band anchor alone (Fig.~1d).

\textbf{Constant count does not predict cost.} \(\mid S \mid\) is not
associated with the minimum attacker budget \(b^{\text{*}}\) across the
six implementations (\(\rho = - 0.563\), \(p = 0.33\)) or the nine
held-out mechanisms for which both are measured (\(\rho = - 0.183\),
exact \(p = 0.638\)), whereas over 400 third-party mechanisms the
association is positive (\(\rho = + 0.483\),
\(p = 8.2 \times 10^{- 25}\)). Because the direction of the association
changes between populations, \(\mid S \mid\) is not a cost model, and we
report \(G\) and \(\mid S \mid\) separately.

\subsection{Rules and judges both fail abruptly, 25$\times$ apart in
attacker
budget}\label{rules-and-judges-both-fail-abruptly-25-apart-in-attacker-budget}

\includegraphics[width=6.1in,height=5.39275in]{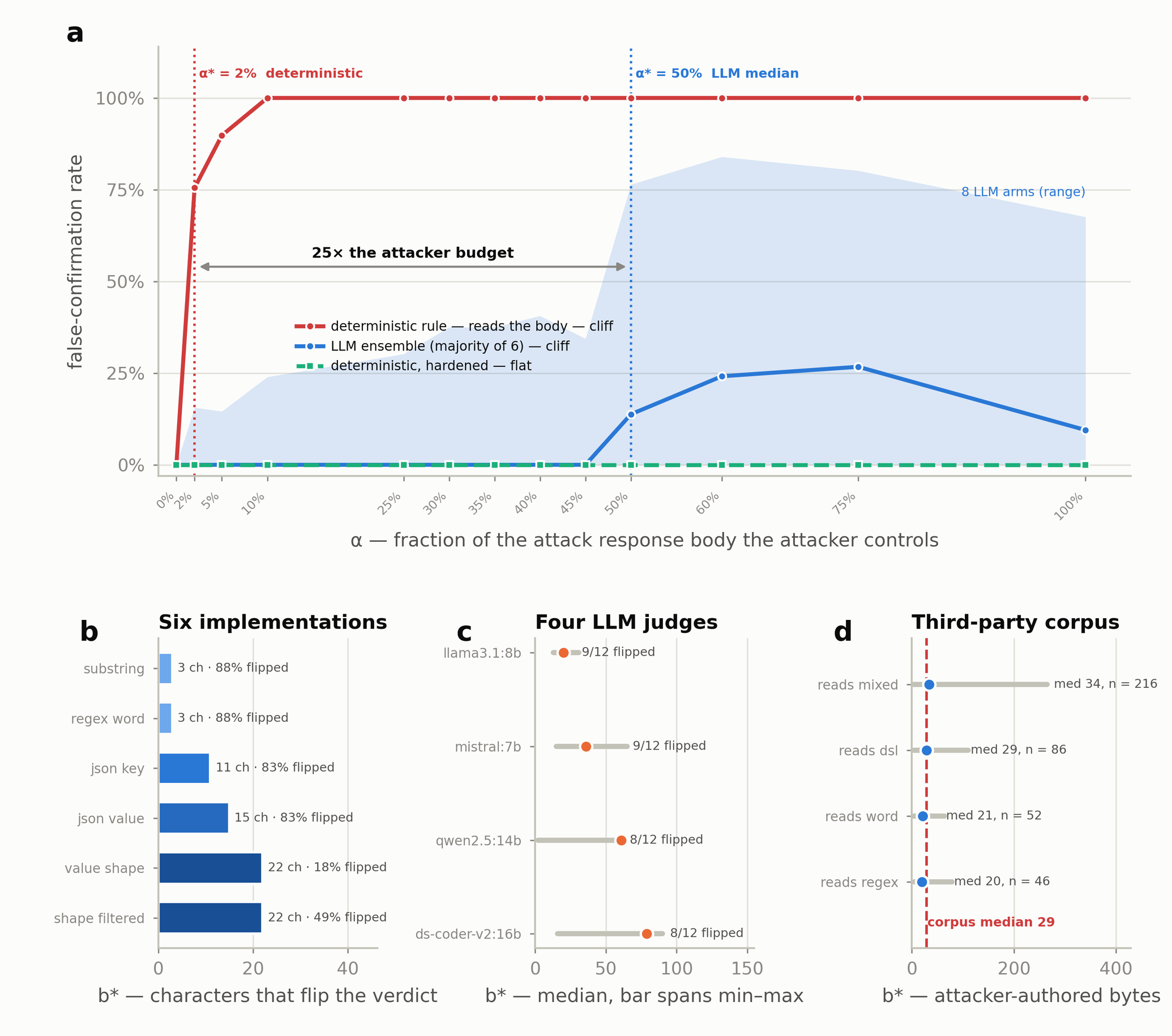}

\textbf{Figure 2. Attacker budget required to forge a confirmation.}
\textbf{(a)} Verifier Robustness Curve: \(F_{A}(\alpha)\) of eq.~(3)
over 147 clean-negative findings and 13 attacker budgets. Red, reachable
deterministic rule; blue, six-judge ensemble; green, hardened rule
(dashed, zero throughout); the shaded band spans the eight model arms.
Dotted verticals mark the breakpoints \(\alpha^{\text{*}}\) of eq.~(4):
2~\% for the rule and a median of 50~\% for the judges, roughly 6 and
157 characters at the median body length. \textbf{(b)} Minimum budget
\(b^{\text{*}}\) of eq.~(6) for six independent implementations of one
check, found by generate-then-minimise search over 161 attackable cases.
Bars give the median in characters; annotations give the share of cases
flipped. Substring and word-boundary regex implementations fall at three
characters (otp, ssn, cvv), a JSON-key check at 11, key-plus-value at 15
and both shape detectors at 22. \textbf{(c)} The same search against
four judges over twelve cases each; dots mark the median
\(b^{\text{*}}\) and bars the range. Judges require a median of 20 to 79
characters, seven to twenty-six times the cheapest rule, and are flipped
on 8--9 of 12 cases. \textbf{(d)} \(b^{\text{*}}\) for 400 sampled
third-party mechanisms, grouped by what the matcher reads; the dashed
line is the corpus median (29 bytes) and the minimum is 3.

A single forgery rate cannot distinguish a verifier that degrades
gradually from one that collapses, and the two call for different
defences; we therefore sweep the attacker\textquotesingle s budget
(Fig.~2a). Both kinds of decider behave as threshold devices: on the
common pool, three of the eight model arms show the same abrupt (cliff)
failure as the rule. What separates them is how much of the response the
attacker must control before the threshold is crossed, a quantity a
defender can budget against: the deterministic rule breaks at
\(\alpha^{\text{*}} = 2\%\) of the response body and the judges at a
median of 50~\%, roughly 6 and 157 characters at the median body length
(Fig.~2a).

Because the curve is a bound for a fixed payload ladder, we also
measured the exact minimum budget \(b^{\text{*}}\) by adaptive search.
Across six independent implementations of one check, \(b^{\text{*}}\)
rises from three characters for substring and word-boundary matching to
11 for a JSON-key check, 15 for key-plus-value and 22 for both shape
detectors (Fig.~2b). Four judges require a median of 20 to 79
characters, seven to twenty-six times the cheapest rule, yet are still
flipped on 8--9 of 12 cases (Fig.~2c). In 400 third-party mechanisms the
median is 29 bytes and the minimum 3 (Fig.~2d), so forgery outside this
pipeline is similarly cheap. The label asserting the strongest warrant
is thus attached to the decision that is cheapest to forge. Per-arm
shape, area, breakpoint and clean false-confirmation rate are given in
Table~S1, and the minimum budgets for every implementation, judge and
third-party matcher class in Table~S2.

\subsection{The result holds across model families; ensemble
statistics depend on the
roster}\label{the-result-holds-across-model-families-ensemble-statistics-depend-on-the-roster}

\includegraphics[width=6.1in,height=5.30435in]{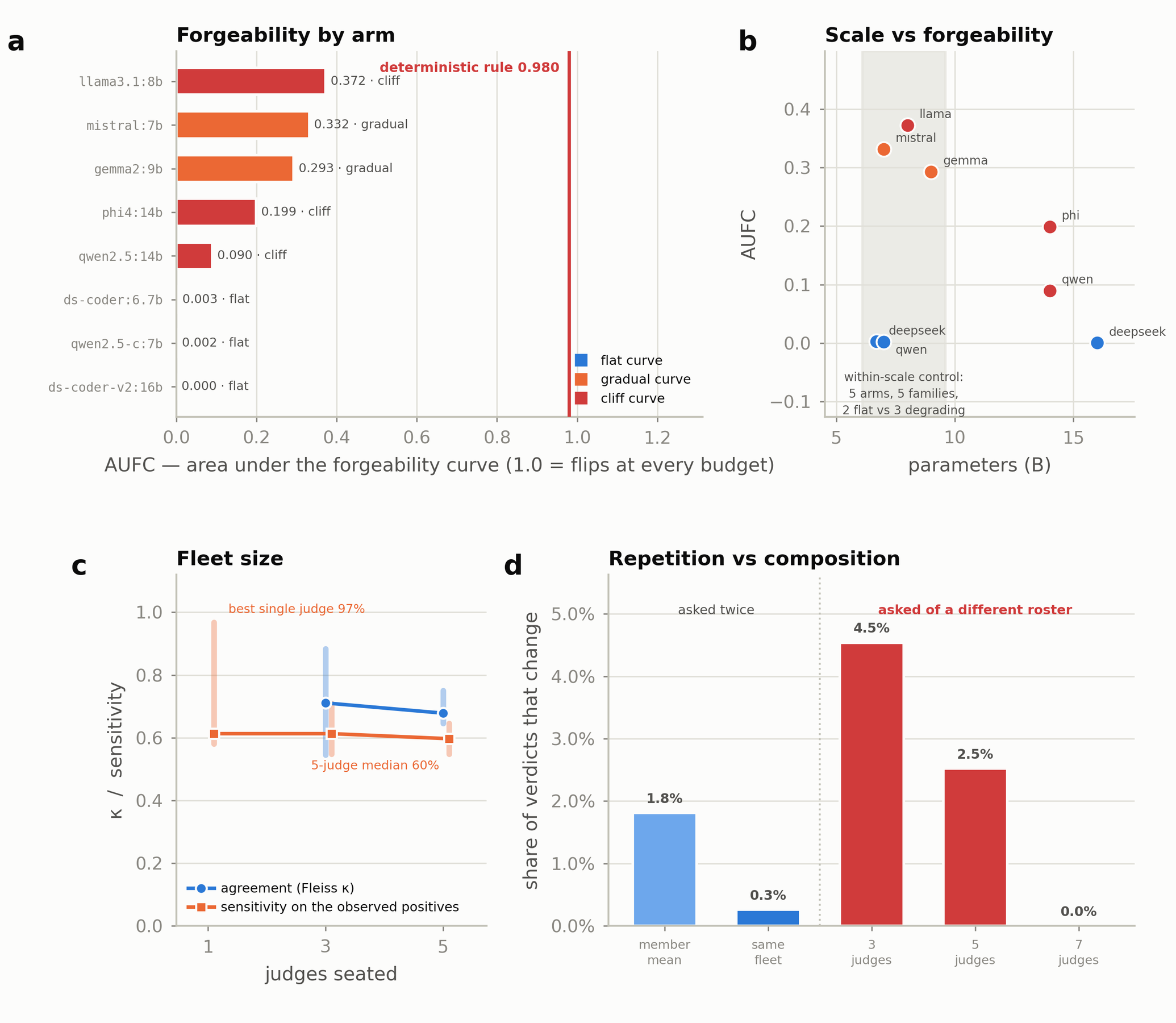}

\textbf{Figure 3. Generality across model families, and variance due to
ensemble composition.} \textbf{(a)} \(AUFC\) of eq.~(4) for the eight
model arms on the 147-case pool, coloured by the shape verdict of
eq.~(5) (blue, flat; orange, gradual; red, cliff). The vertical line
marks the reachable deterministic rule (0.980); the nearest model arm is
0.607 below it, and the hardened rule scores 0.000. \textbf{(b)} AUFC
against parameter count, labelled by family, with the 6.7--9B band
shaded (\(\rho = - 0.193\), exact \(p = 0.654\)). Within that band, five
arms from five families split two flat against three degrading.
\textbf{(c)} Fleiss \(\kappa\) over the 39 observed findings (blue) and
sensitivity on the 31 findings judged by every arm (orange) against the
number of judges, computed on the two six-judge rosters described in
Methods. Dots are medians over rosters of each size and bars the range.
At three judges, \(\kappa\) ranges from \(+ 0.545\) to \(+ 0.883\) with
only membership varying; median sensitivity does not increase with size
(best single judge 97~\%, six-judge fleet 55~\%). Values are listed in
Tables~S6 and S8. \textbf{(d)} Two sources of verdict variation on the
same 34 cases. \emph{Left}: repeated queries to the same judge (ten
times) or the same fleet (twice) change 1.8~\% and 0.3~\% of verdicts.
\emph{Right}: a differently composed fleet changes 4.5~\% of (roster,
case) verdicts at three judges and 2.5~\% at five; at seven judges only
one roster exists.

To test generality, we evaluated eight model arms from six families on a
common pool. Three arms fail as cliffs, but the most vulnerable still
has an area under the forgeability curve 0.607 below that of the
deterministic rule (0.980), and the hardened rule scores 0.000
(Fig.~3a). Robustness does not track parameter count
(\(\rho = - 0.193\), exact \(p = 0.654\); Fig.~3b); within the 6.7--9B
band, a small model from a robust family stays flat while a small model
from a less robust family degrades. Scales above 16B and sampled
decoding were not evaluated on this pool.

Judges are highly self-consistent: across 2,380 inferences (seven
judges, ten repetitions of identical evidence), per-judge
self-consistency is 94--100~\% and fleet self-consistency 99.7~\%, so
disagreement between judges cannot be attributed to sampling noise.
Fleet composition, however, changes verdicts. With three judges, Fleiss
\(\kappa\) ranges from \(+ 0.545\) to \(+ 0.883\) depending only on
which judges are seated, and median sensitivity does not increase with
fleet size: the best single judge confirms 97~\% of observed positives
and the six-judge fleet 55~\% (Fig.~3c). Replacing members changes the
verdict on 6 of 34 cases, whereas repeating the same judge or fleet
changes 1.8~\% and 0.3~\% of verdicts (Fig.~3d), a source of variance
that repeatability checks cannot detect because the evidence is
unchanged. Ensemble statistics for every fleet size and roster are
listed in Table~S6.

Within the six-judge agreement fleet, agreement is substantial (Fleiss
\(\kappa = + 0.683\), Krippendorff \(\alpha = + 0.685\)) and bimodal,
with 74~\% of findings at the extremes of the vote (Fig.~S1a). It is not
organised by training lineage: pairs sharing a lineage agree only
\(+ 0.040\) more than other pairs (exact permutation \(p = 0.30\);
Fig.~S1b), whereas removing a single judge changes fleet \(\kappa\) by
up to \(+ 0.066\) (Fig.~S1c). Per-judge centrality, leave-one-out
effect, self-consistency and cost are given in Table~S8.

\subsection{No implementation of one check is both attack-resistant
and
precise}\label{no-implementation-of-one-check-is-both-attack-resistant-and-precise}

\includegraphics[width=6.1in,height=5.30435in]{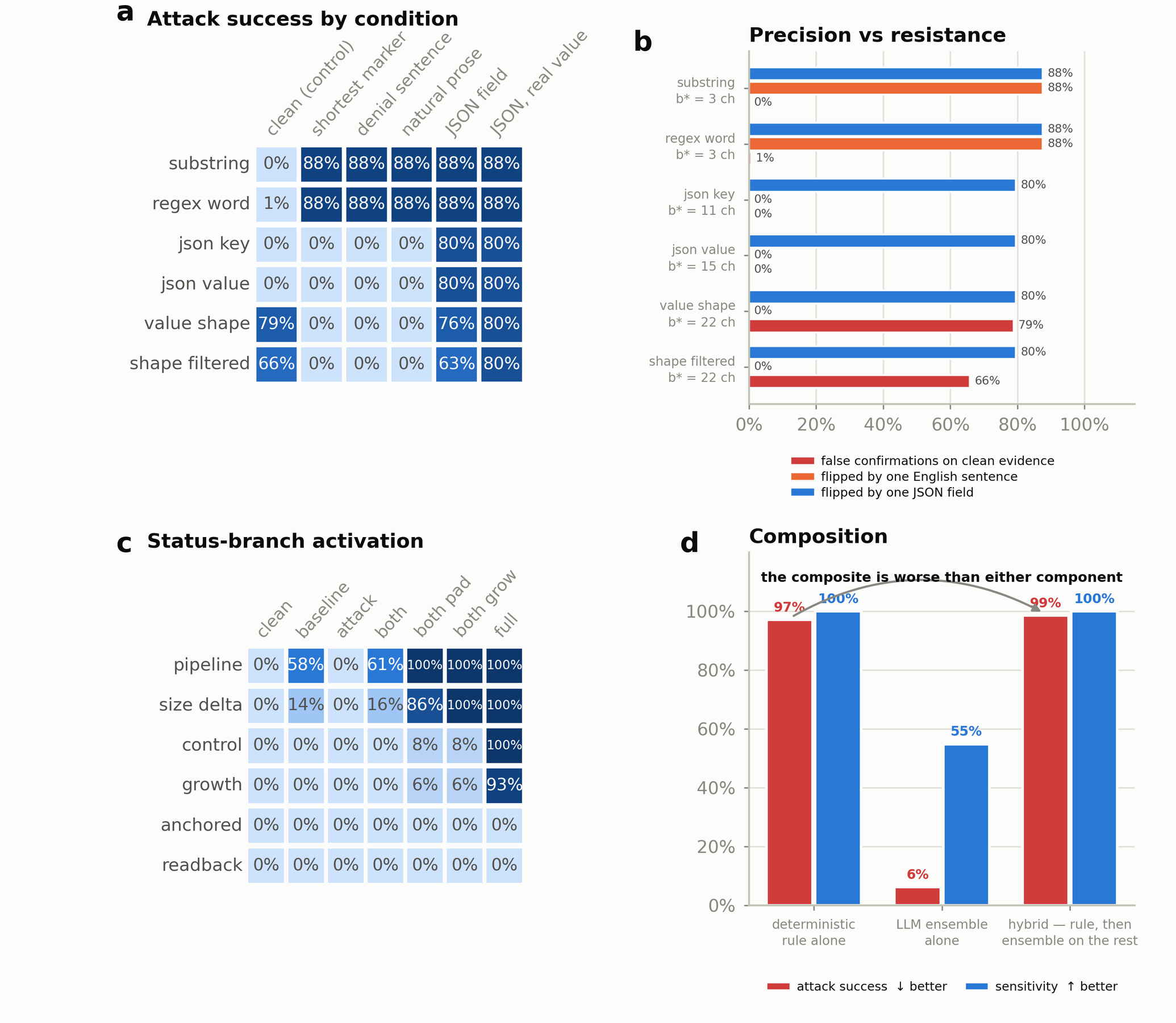}

\textbf{Figure 4. Six implementations of one check, and their
composition with a judge ensemble.} \textbf{(a)} Attack success for six
independent implementations of confirm\_unauth (rows) under six
conditions (columns): the share of 161 attackable cases flipped. The
clean control column is non-zero because the two value-shape
implementations fire on 79~\% and 66~\% of unmodified evidence.
Text-matching implementations fall to a single English sentence,
structure-parsing ones to a single injected JSON field, and all six to a
high-entropy JSON value. \textbf{(b)} The same implementations on three
rates, with \(b^{\text{*}}\) in the row label: clean false-confirmation
rate (red), share flipped by one English sentence (orange) and by one
JSON field (blue). The two implementations that resist prose are the two
with high clean false-confirmation rates. \textbf{(c)} The
status-inversion branch over 200 instances: six implementations (rows)
under activation conditions ranging from clean evidence to three
attacker-authored fields (columns). The shipped rule returns
inconclusive on all unmodified instances through both branches; one
authored status code activates its three-way conjunction on 58~\% of the
pool, and three fields on 100~\%. The anchored and read-back
implementations never activate. \textbf{(d)} The rule-then-ensemble
hybrid against its components, on the 31 observed positives with
verdicts from every arm. Red, attack success; blue, sensitivity. The
hybrid confirms 100~\% and is forged on 99~\%, more than the
deterministic rule (97~\%) or the ensemble (6~\%), because it confirms
when either component confirms.

The SQL-injection confirmer returns inconclusive on every real finding
in this corpus, which consists of authorisation evidence, so forging it
would only demonstrate activation of a dormant rule; this is why the
marker rule flips on 100~\% of cases in our initial master-key test
while the judges flip on at most 20~\% (Fig.~S1d). We therefore attacked
a mechanism that fires in practice, confirm\_unauth, and implemented it
six independent ways. It fires on 39 of 200 instances and declines on
the other 161, which form the attack set \(N_{A}\). The full grid is
given in Table~S3 and the status-inversion branch in Table~S4.

Appending the sentence \emph{"no password is returned to the client"},
which tells a human reader that the endpoint is safe, flips both
text-matching implementations on 88~\% of cases (Fig.~4a). Because a
benign API could return such a sentence, the same defect also produces
false positives in normal operation. Parsing resists prose but falls to
one injected field (80~\%), and every implementation falls to a
high-entropy JSON value (Fig.~4a). Shape detection resists prose, but
fires on two-thirds of unmodified traffic, because response bodies
contain UUIDs and digests with the length, character mix and entropy of
a credential; explicitly excluding those formats reduces the clean
firing rate only from 79~\% to 66~\%. The two implementations that
resist prose are therefore exactly the two with high clean
false-confirmation rates, and none is low on all three rates (Fig.~4b).
The weakness therefore lies in the task definition: detecting
unauthenticated exposure of sensitive data from a response body requires
a decision made from attacker-authored data, whatever the
implementation.

The same holds for the status-inversion branch of the SQL-injection
confirmer. The shipped rule returns inconclusive on all 200 unmodified
instances, but one attacker-authored status code satisfies its three-way
conjunction on 58~\% of the pool and three authored fields on 100~\%,
whereas the anchored and read-back implementations cannot be activated
at all (Fig.~4c; Table~S4). A conjunction costs the attacker only the
conjuncts that are not already true.

Composition does not help. The rule-then-ensemble hybrid confirms 100~\%
of observed positives and is forged on 99~\% of attacked evidence, more
than either the deterministic rule (97~\%) or the ensemble (6~\%),
because it confirms whenever either component does (Fig.~4d). Hardening
the deterministic component does not repair the hybrid, because the
hardened rule confirms none of these findings and the hybrid then
reduces to the ensemble. Only conjunctive routing improves robustness,
and on this corpus it confirms nothing.

\textbf{End-to-end demonstration.} To confirm that the offline results
transfer to a live system, we served attacker-authored responses from an
endpoint on the loopback interface, fetched them with the
pipeline\textquotesingle s own HTTP client over TCP, and applied the
pipeline\textquotesingle s own confirmation rule. Three of five
responses, none of which disclosed any data, were confirmed as
vulnerabilities and would have reached an operator labelled
\emph{confirmed by observation}.

\subsection{Moving the decisive value off the
attacker\textquotesingle s channel closes the attack, at a
cost}\label{moving-the-decisive-value-off-the-attackers-channel-closes-the-attack-at-a-cost}

\includegraphics[width=6.1in,height=5.83478in]{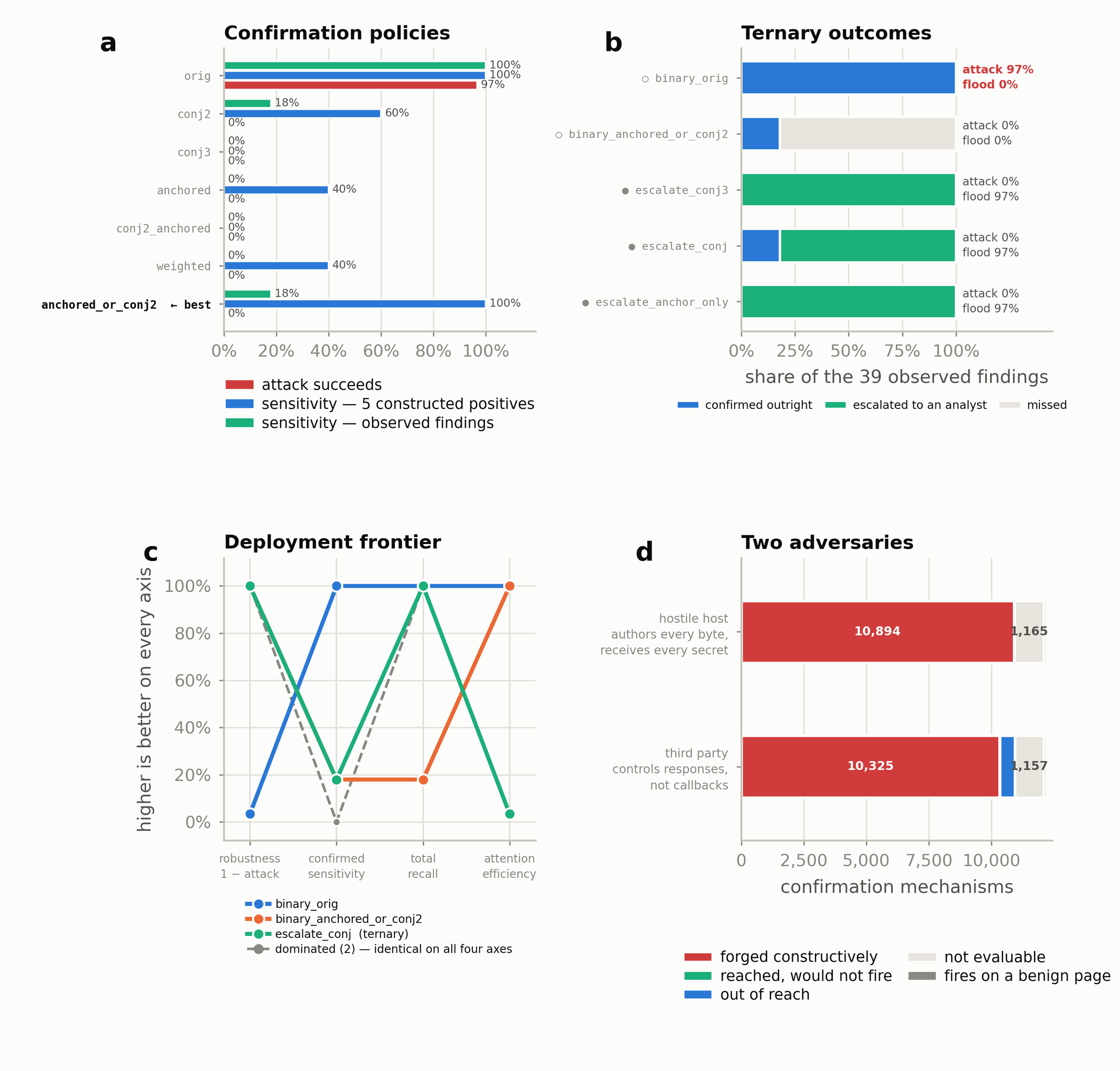}

\textbf{Figure 5. The defence, its cost, and the adversary against which
it fails.} \textbf{(a)} Seven confirmation policies over 1,932 attacked
negatives. Red, attack success; blue, sensitivity on five constructed
positives; green, sensitivity on the 39 observed findings.
anchored\_or\_conj2 (eq.~(7)) reduces attack success from 97~\% to 0~\%
with unchanged repeatability and cost. On constructed positives its
sensitivity is unchanged; on observed findings it falls from 100~\% to
18~\%, because each observed finding fires a single mechanism and the
policy requires corroboration. \textbf{(b)} Outcome shares of the
observed positives under each policy: confirmed (blue), escalated to an
analyst (green) and missed (grey); filled circles mark ternary and open
circles binary policies, with attack success and escalation rate under
attack annotated. escalate\_conj (eq.~(8)) makes the same confirmation
decisions as the hardened binary policy and escalates the remainder:
100~\% total recall, 0~\% missed and 0~\% attack success, with
escalation of 97~\% of attacked evidence and 0~\% of clean evidence.
\textbf{(c)} Five policies on the four higher-is-better axes defined in
Methods, shown as parallel coordinates because three policies coincide
on any two axes. Coloured lines are the three frontier policies; the
grey dashed line is the two dominated policies, which coincide.
binary\_orig is on the frontier only through its unattacked sensitivity,
and is forged by a three-character payload. \textbf{(d)} The third-party
corpus under two adversaries. \emph{Bottom}: a third party that controls
responses but not callbacks. \emph{Top}: the scanned host, which
receives the interactsh\textsuperscript{60} callback URL in the payload
and the per-run value in the request; a further 569 mechanisms fall and
none remains out of reach.

The criterion identifies the remedy. A verifier is forgeable to the
extent that its decision reads attacker-controlled data, so hardening
means moving the decisive value to a channel the attacker cannot write
(eq.~(7)), not adding determinism. The pipeline already contains such
anchors, namely the mechanisms with \(G = 0\): a per-run nonce reflected
back, an out-of-band canary callback, and a sentinel value that the
server persisted. Where a vulnerability class has such an anchor, it
should be used; where it has none, confirmation should require
corroboration or be referred to a human. The policy anchored\_or\_conj2
(eq.~(7)) reduces attack success over 1,932 attacked negatives from
97~\% to 0~\% with unchanged repeatability and cost (Fig.~5a). Its
sensitivity is unchanged on five constructed positives but falls from
100~\% to 18~\% on the 39 observed findings, because each observed
finding fires a single mechanism and the policy requires corroboration
(Fig.~5a). All policies are scored in Table~S5, and Table~3 compares the
resulting systems on six axes.

\textbf{Table 3. Verifier systems compared on six axes.} Sensitivity is
measured on the 31 findings the pipeline labelled positive. Attack
success is the mean false-confirmation rate over the 12 non-zero
attacker budgets of the $\alpha$-sweep (147 cases); it is therefore not the
AUFC of Table S1.

\includegraphics[width=6.1in,height=2.24721in]{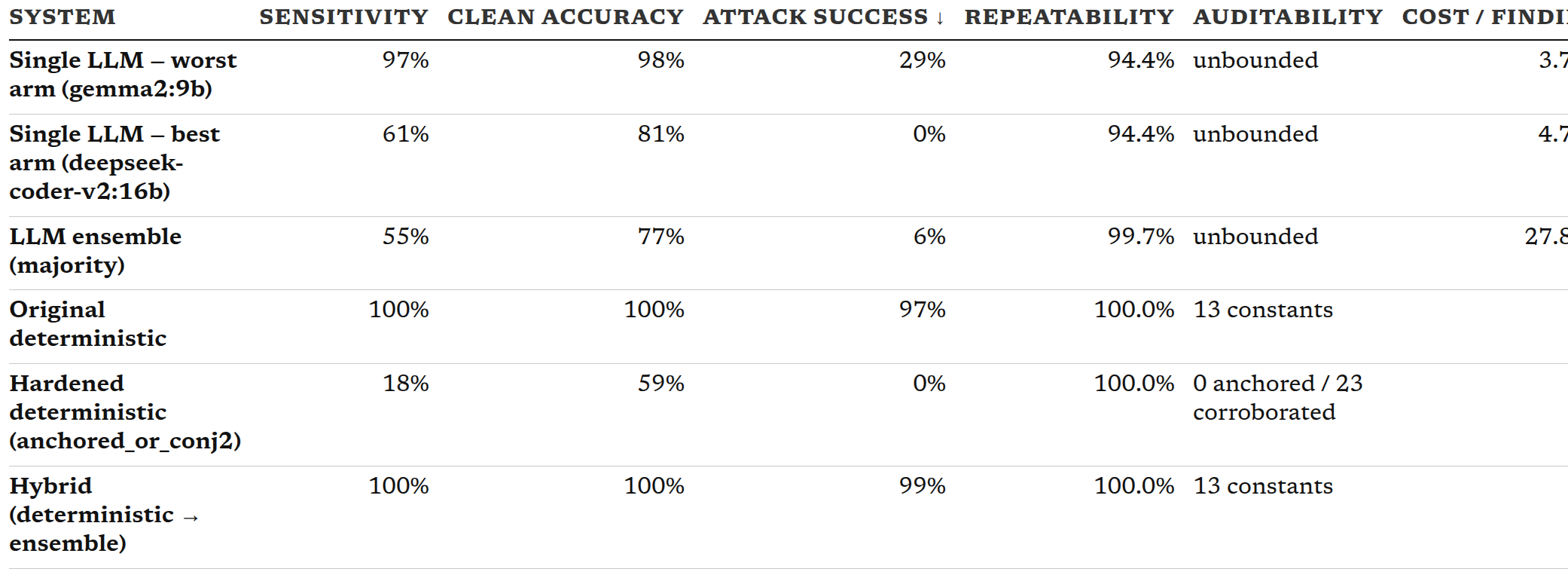}

Note. On five hand-constructed positives (one per signal type) the
original and hardened rules both score 100\%; on the observed positives
they score 100\% and 18\%, because each observed finding fires a single
mechanism and the hardened policy requires corroboration. Auditability
is the number of constants the decision consults: for the original rule,
the marker list of Table 1; for the hardened policy, none on its
anchored path and the reachable mechanisms\textquotesingle{} constants
on its corroboration path (Table S5). Model arms have no enumerable
list.

The cost in sensitivity is substantial. A conjunctive policy is
affordable only where findings produce corroborating signals; where they
produce a single signal, the binary choice is between an attackable
confirmation and an inconclusive verdict referred to a human. A third
output (eq.~(8)), which escalates such findings together with the probe
that would resolve them, keeps attack success at 0~\% with 100~\% total
recall and no missed positives; it escalates 97~\% of attacked evidence
and none of the clean evidence (Fig.~5b). Escalations then occur only
under attack, which is the appropriate failure direction for a triage
system. On the four scoring axes defined in Methods, three policies form
the frontier (Fig.~5c): the hardened binary policy, its ternary
counterpart, and the original rule, which is on the frontier only
through its unattacked sensitivity and is forged by a three-character
payload.

These results assume a third-party adversary. The scanned host itself
receives the out-of-band callback URL in the payload and the per-run
value in the request; under that adversary a further 569 third-party
mechanisms are forged and none remains out of reach (Fig.~5d).

\section{Discussion}\label{discussion}

The distinction between \emph{judged} and \emph{confirmed} findings is
real but does not follow the stochastic/deterministic divide. What
matters is whether the decisive value is on a channel the attacker
writes; by that criterion the rule and the judge fall on the same side
and differ in threshold rather than in kind. Three practical
recommendations follow.

\textbf{Audit the channel, not the paradigm.} Whether a decision reads
attacker-controlled data can be determined from source code before any
attack is attempted: the gate predicted all 16 held-out mechanisms, and
none of the 10,925 scored third-party mechanisms was predicted
unreachable and then forged (Fig.~1b). The number of constants a
decision consults is not a substitute: it neither separates forgeable
from unforgeable mechanisms (Fig.~1a) nor orders their cost consistently
(Fig.~2b,d). This makes the check suitable for routine code review.

\textbf{Report the roster with the number.} Every ensemble statistic in
this study changed with fleet composition on identical evidence
(Fig.~3c,d), and a single judge moved fleet agreement more than training
lineage did (Fig.~S1b,c); repeatability checks detect neither effect.
Agreement, sensitivity or confidence reported for a panel of models
should therefore be accompanied by the panel\textquotesingle s exact
membership.

\textbf{Prefer a ternary verdict.} Every hardening we measured traded
sensitivity for robustness until an \emph{escalate} verdict was allowed
(Fig.~5a,b). Neither a different implementation (Fig.~4a,b) nor
composition with a judge ensemble (Fig.~4d) removed the trade-off. With
it, the trade-off disappears, at the cost of analyst attention only when
an attack is under way.

The findings also bear on evaluation. Benchmarks for offensive and
defensive agents often decide success by matching a string in the
output\textsuperscript{10,11}. Such criteria are confirmation mechanisms
with \(G = 1\): they can be satisfied without performing the task, and
an agent optimised against them is optimised against a proxy that can
diverge from the intended goal\textsuperscript{53,55}. Our judges needed
only 20 to 79 characters to be flipped (Fig.~2c), and a string-matching
criterion only three (Fig.~2b). A benchmark whose success signal the
system under test can author does not measure capability reliably.

\subsection{Limitations}\label{limitations}

\textbf{One pipeline and evidence distribution.} The evidence comprises
200 instances from five target families, 84~\% from one application. The
six re-implementations and the third-party corpus show that the weakness
is not specific to one codebase, but all reported rates are properties
of this distribution, and the third-party corpus was scored from matcher
specifications rather than executed against live hosts.

\textbf{Open-weight judges only.} The eight model arms are 6.7--16B
open-weight models, run locally with greedy decoding. We make no claim
about hosted frontier models, to which hard cases are most likely to be
sent, or, on this pool, about larger models or sampled decoding.
Evaluating hosted judges is the most important extension.

\textbf{Mechanism-derived labels.} The positive class is labelled by the
pipeline\textquotesingle s own mechanisms, which this paper shows to be
forgeable. These mechanisms return inconclusive rather than guess, which
supports but does not establish their labels. Fifty blinded annotation
packets, stratified and stripped of every field stating a pipeline
conclusion, have been prepared for two expert annotators; of the four
steps of the labelling procedure, only the first is complete (Fig.~S2a);
a machine-annotated pilot on the same packets, reported as a pilot
rather than ground truth, finds that the implementations most resistant
to forgery agree least with a blinded reader (73~\% agreement for
substring matching against 34~\% and 23~\% for the two shape detectors;
Fig.~S2b,c).

\textbf{No operator data.} We measure the machine, not the analyst. We
have designed an operator study in which analysts see the same finding
under labels ranging from a hedge to an assertion of an observed state
change (Fig.~S3a). It cannot yet be run: the run records contain one
matched pair against a pre-registered minimum of eight (Fig.~S3b), and
every candidate stimulus for two of the three studies lies on a publicly
taught target, so provenance cannot be separated from recognition
(Fig.~S3c). Power simulations indicate 24--48 participants for the
binary contrast at effect sizes of 0.4--0.8 log-odds (Fig.~S3d), but
show that a two-level design cannot distinguish a gradual effect from a
step (Fig.~S3e). Design targets are listed in Table~S9.

\textbf{Bounds.} Each point on a robustness curve is a lower bound on
attack success, because the attacker draws from a fixed payload list;
each \(b^{\text{*}}\) is an upper bound on the minimum attacker cost.
Where the two disagree, \(b^{\text{*}}\) is the tighter statement.

\textbf{Adversary dependence.} The unreachable mechanisms and the
hardened policy rely on anchors the attacker cannot reach, and
reachability depends on the adversary. A scanned host receives the
out-of-band callback URL in the payload and the per-run value in the
request; against that adversary none of the third-party mechanisms
remains out of reach (Fig.~5d). Bug-bounty triage bots and agentic
scanners operate under this model. All claims of unforgeability in this
paper hold against a third party, not against a hostile target.

\textbf{Live demonstration.} The end-to-end test shows that the pipeline
confirms attacker-authored responses fetched over HTTP; it does not show
how often an attacker attains that position.

\section{Ethics and disclosure}\label{ethics-and-disclosure}

No third-party system was scanned, contacted or attacked. The live
demonstration used a loopback endpoint serving invented data. The
third-party corpus is a public, openly licensed collection of scanner
templates\textsuperscript{59}; forgeries were constructed against
matcher specifications, never against hosts, and no template author is
named in connection with a defect. The vulnerabilities in the evidence
corpus are in deliberately vulnerable applications maintained for
security education. The weakness is a design property of a class of
confirmation mechanisms rather than a flaw in a single product; because
the attack requires only a few characters and no tooling, we publish it
together with the criterion and the mitigation.

\section{Methods}\label{methods}

\subsection{Pipeline and evidence}\label{pipeline-and-evidence}

The pipeline is a four-stage API-security pipeline
(generate~$\rightarrow$~judge~$\rightarrow$~confirm~$\rightarrow$~measure) in which the third stage is the
verdict authority: a finding is accepted when a deterministic mechanism
observes the target enter a forbidden state. The evidence consists of
457 distinct instances from live scans of five target families; analyses
use a 200-instance stratified pool and, for the robustness curves, the
147 instances on which every arm is a clean negative. 84~\% of the pool
comes from one application. The \emph{observed positive class} is the 39
findings that the pipeline\textquotesingle s own mechanisms confirm on
unmodified evidence, and policies are scored against all 39. Comparisons
involving model arms (Fig.~3c, Fig.~4d, Table~3) use the 31 of these
that carry a verdict from every arm; the remaining 8 have no model
verdict and are excluded rather than imputed.

\subsection{Judges}\label{judges}

Eleven open-weight models are used across four rosters, and each
ensemble statistic is reported with its roster. The robustness curves
use \textbf{eight arms} from six families (qwen, deepseek, llama, gemma,
mistral, phi) at 6.7--16B. Agreement, the lineage contrast, the
leave-one-out analysis and the master-key grid use a \textbf{six-judge
agreement fleet}: qwen2.5:14b, qwen2.5:32b, qwen2.5-coder:32b,
deepseek-r1:32b, phi4:14b and deepseek-coder-v2:16b. The sensitivity
sweep uses a \textbf{six-judge sweep fleet} in which llama3.1:8b,
mistral:7b and gemma2:9b replace the three models above 16B, and the
repeatability study uses these six plus deepseek-r1:32b. All models are
greedy-decoded and served locally through Ollama on one RTX~5090; no
hosted model is used.

\subsection{Threat model}\label{threat-model}

The attacker controls \textbf{the body an endpoint returns} and may
choose its status code and headers. This capability arises from
server-side injection, a compromised upstream service, an
attacker-registered tenant on a multi-tenant API, or server-side request
forgery that directs a scanner to an attacker-owned host. The attacker
does \emph{not} control values the tester generates per run, out-of-band
callbacks to tester infrastructure, or what the real server persisted
and returns on an independent read-back. The Limitations discuss the
case in which the last assumption fails.

\subsection{The auditable attack
surface}\label{the-auditable-attack-surface}

Let \(D\) be a decision procedure returning a confirmation verdict,
\(S(D)\) the set of literal constants it consults (marker lists, status
sets, regular expressions, thresholds), and let

\begin{equation}
\begin{matrix}
R(D) = \mathbb{1}\lbrack\text{attacker-controlled input reaches }D\rbrack, \\
K(D) = \mathbb{1}\lbrack\text{the decisive value is attacker-knowable}\rbrack.
\end{matrix}
\end{equation}
The \emph{auditable attack surface} is the constant count gated by
reachability:

\begin{equation}
AAS(D) = \left\{ \begin{matrix}
 \mid S(D) \mid , & R(D) \cdot K(D) = 1, \\
0, & \text{otherwise.}
\end{matrix} \right.\
\end{equation}
AAS is not a probability. It counts what an auditor must read to
determine how a decision can be driven, and is zero when no
attacker-reachable input affects it. An LLM judge has \(R = K = 1\) and
an unbounded, non-enumerable \(\mid S \mid\). We call
\(G(D) = R(D)K(D)\) the \emph{reachability gate} and report it
separately from \(\mid S \mid\), because the gate predicts forgeability
across populations and the magnitude does not. A reachable mechanism
with no literal constant (\(G = 1\), \(\mid S \mid = 0\); one case in
Table~2) has \(AAS = 0\) but is classified by \(G\), which is the
quantity used for prediction.

\subsection{Attacker budget and the robustness
curve}\label{attacker-budget-and-the-robustness-curve}

Let \(x\) be an evidence instance with response body \(b(x)\), let
\(x \oplus p\) denote \(x\) with payload \(p\) composed into that body,
and let \(\mathcal{P}\) be the payload families available to the
attacker. For an arm \(A\), let \(N_{A} = \{ x:A(x) = \text{decline}\}\)
be the instances on which it is correct before the attack. The budget
\(\alpha \in \lbrack 0,1\rbrack\) is the fraction of the body the
attacker may author, and the \emph{false-confirmation rate} is

\begin{equation}
F_{A}(\alpha) = \frac{1}{\mid N_{A} \mid}\sum_{x \in N_{A}}^{}\mathbb{1}\lbrack\ \exists\, p \in \mathcal{P}:\  \mid p \mid \leq \alpha \mid b(x) \mid ,A(x \oplus p) = \text{confirm}\ \rbrack.
\end{equation}
Restricting to \(N_{A}\) is the clean-accuracy control: no forgery is
scored against a detector that was already wrong. Over a grid
\(\alpha_{1} < \ldots < \alpha_{m}\) we summarise the curve by its
normalised area and breakpoint,

\begin{equation}
\begin{matrix}
AUFC(A) = \frac{1}{\alpha_{m} - \alpha_{1}}\int_{\alpha_{1}}^{\alpha_{m}}F_{A}(\alpha)\, d\alpha, \\
\alpha_{A}^{\text{*}} = \min\{\alpha:F_{A}(\alpha) \geq F_{A}^{\min} + \frac{1}{2}\rho_{A}\},
\end{matrix}
\end{equation}
with rise \(\rho_{A} = F_{A}^{\max} - F_{A}^{\min}\) and the integral
evaluated by the trapezoid rule. A shape verdict is issued only when the
rise exceeds a noise floor that scales with the sample size behind each
point,

\begin{equation}
\varepsilon_{A} = \max\left( 0.05,\mspace{6mu} 2\sqrt{{\bar{F}}_{A}\left( 1 - {\bar{F}}_{A} \right)/n},\mspace{6mu} 2/n \right),
\end{equation}
so that at small \(n\) a change in one or two findings is not read as
degradation. Where \(\rho_{A} > \varepsilon_{A}\), the arm is a
\emph{cliff} if its 10--90~\% transition width
\(W_{A} = \left( \alpha_{90} - \alpha_{10} \right)/\left( \alpha_{m} - \alpha_{1} \right)\)
satisfies \(W_{A} \leq 0.25\) and \emph{gradual} otherwise; when the
transition falls within a single wide grid interval, the estimator
returns unresolved. The grid is dense below \(\alpha = 10\%\) and
refined at 30, 35, 40, 45 and 60~\%, so that shape verdicts are not
artefacts of grid resolution.

Because the curve is a bound for a laddered adversary, we also compute
the exact \emph{minimum budget} in characters by generate-then-minimise
search:

\begin{equation}
b_{A}^{\text{*}}(x) = \min\{ \mid p \mid :\ p \in \mathcal{P},\ A(x \oplus p) = \text{confirm}\},
\end{equation}
reported per arm as the median over \(N_{A}\). Each \(b^{\text{*}}\) is
an upper bound on the true minimum.

\subsection{Confirmation policies and
scoring}\label{confirmation-policies-and-scoring}

Let \(M = U\,\dot{\cup}\, R\) partition the mechanisms into unreachable
(\(G = 0\)) and reachable (\(G = 1\)) ones, and let
\(m(x) \in \{ 0,1\}\) indicate whether \(m\) fires on \(x\). The
hardened binary policy is

\begin{equation}
\begin{matrix}
\Pi_{\text{hard}}(x) = \text{confirm} \Leftrightarrow \ \left( \bigvee_{m \in U}m(x) \right) \\
 \vee \ \left( \sum_{m \in R}^{}m(x)\  \geq 2 \right),
\end{matrix}
\end{equation}
and its ternary counterpart escalates what neither branch decides:

\begin{equation}
\Pi_{\text{esc}}(x) = \left\{ \begin{matrix}
\text{confirm}, & \Pi_{\text{hard}}(x) = \text{confirm}, \\
\text{escalate}, & \exists m \in M:m(x) = 1, \\
\text{decline}, & \text{otherwise.}
\end{matrix} \right.\
\end{equation}
Over an observed positive class \(P\), an attacked negative set
\(N^{\dagger}\) and a clean negative set \(N\), we score four
higher-is-better axes: robustness
\(1 - \Pr_{N^{\dagger}}\lbrack\text{confirm}\rbrack\), confirmed
sensitivity \(\Pr_{P}\lbrack\text{confirm}\rbrack\), total recall
\(\Pr_{P}\lbrack\text{confirm} \vee \text{escalate}\rbrack\), and
attention efficiency
\(1 - \Pr_{N^{\dagger}}\lbrack\text{escalate}\rbrack\). A policy \(\Pi\)
is dominated if some \(\Pi'\) is at least as good on all four axes
and strictly better on one. Both sensitivity axes are needed, because
confirmed sensitivity alone treats an escalated positive as a missed
one.

\subsection{Controls}\label{controls}

We applied three controls. \emph{Clean accuracy}: eq.~(3) restricts
every attack to \(N_{A}\). \emph{Deletion only}: truncating evidence
without adding content flips 0 of 161 cases on every arm, so the curves
measure forgery rather than loss of evidence. \emph{Dormancy}: a rule
that returns inconclusive on every real instance is trivially flipped,
so the six-implementation study attacks a mechanism that fires on 39 of
200 instances.

\subsection{Statistics and
pre-registration}\label{statistics-and-pre-registration}

We use exact permutation tests and Fisher\textquotesingle s exact test,
because the panels are small; where a design has a smallest attainable
\(p\), it is reported. Spearman correlations over the 400-mechanism
corpus sample use the \(t\) approximation. For the held-out test, every
prediction was read from source and hashed before any attack was written
(PREDICTION\_SHA~=~a41c7a1fec100b9f).

\subsection{Reproducibility and
compute}\label{reproducibility-and-compute}

All judging ran locally through Ollama on one RTX~5090, with no external
API, for approximately 16 GPU-hours in total. Deterministic audits are
exhaustive, with no sampling. Every figure and table is generated
programmatically from the machine-written record of the analysis that
produced it.

\subsection{Author contributions}\label{author-contributions}

A.F. and N.S. designed the study. A.F. built the pipeline, the attack
apparatus and the analysis, and ran the experiments. N.S. designed the
audit criterion and the held-out prediction protocol. Both authors
analysed the results and wrote the manuscript.

\subsection{Data and code
availability}\label{data-and-code-availability}

The analysis records from which every figure and table is generated, the
vector figures and the bibliography are provided with this preprint. The
evidence corpus is drawn from deliberately vulnerable applications
maintained for security education; the third-party template corpus is
public and openly licensed\textsuperscript{59} and is not redistributed.
The code used in this study is available from the corresponding author
on reasonable request. Because parts of it implement working attacks on
security verification mechanisms, access to those components may be
limited, or granted after review of the intended use. No human-subject
data were collected.

\subsection{Competing interests}\label{competing-interests}

N.S. is employed by OmiCore Inc., which develops the security pipeline
audited here. A.F. is a graduate student at Kyushu University and
contributes to OmiCore Inc. as an unpaid volunteer. All defects found in
that pipeline are reported, together with the mitigation.

\section{References}\label{references}

\begin{enumerate}
\def\labelenumi{\arabic{enumi}.}
\tightlist
\item
  B. A. Alahmadi, L. Axon and I. Martinovic. \emph{99\% False Positives:
  A Qualitative Study of SOC Analysts\textquotesingle{} Perspectives on
  Security Alarms.} 31st USENIX Security Symposium, 2022.
\item
  B. Johnson, Y. Song, E. Murphy-Hill and R. Bowdidge. \emph{Why
  Don\textquotesingle t Software Developers Use Static Analysis Tools to
  Find Bugs?.} ICSE, 2013. doi:10.1109/ICSE.2013.6606613.
\item
  M. Christakis and C. Bird. \emph{What Developers Want and Need from
  Program Analysis: An Empirical Study.} ASE, 2016.
  doi:10.1145/2970276.2970347.
\item
  C. Sadowski, E. Aftandilian, A. Eagle, L. Miller-Cushon and C. Jaspan.
  \emph{Lessons from Building Static Analysis Tools at Google.}
  Communications of the ACM 61(4):58--66, 2018. doi:10.1145/3188720.
\item
  J. Smith, L. N. Q. Do and E. Murphy-Hill. \emph{Why
  Can\textquotesingle t Johnny Fix Vulnerabilities: A Usability
  Evaluation of Static Analysis Tools for Security.} SOUPS, 2020.
\item
  A. Doupé, M. Cova and G. Vigna. \emph{Why Johnny Can\textquotesingle t
  Pentest: An Analysis of Black-box Web Vulnerability Scanners.} DIMVA,
  2010. doi:10.1007/978-3-642-14215-4\_7.
\item
  J. Bau, E. Bursztein, D. Gupta and J. Mitchell. \emph{State of the
  Art: Automated Black-Box Web Application Vulnerability Testing.} IEEE
  Symposium on Security and Privacy, 2010. doi:10.1109/SP.2010.27.
\item
  R. Fang, R. Bindu, A. Gupta and D. Kang. \emph{LLM Agents can
  Autonomously Exploit One-day Vulnerabilities.} arXiv preprint, 2024.
  arXiv:2404.08144.
\item
  G. Deng, Y. Liu, V. Mayoral-Vilches, P. Liu, Y. Li, Y. Xu, T. Zhang,
  Y. Liu, M. Pinzger and S. Rass. \emph{PentestGPT: Evaluating and
  Harnessing Large Language Models for Automated Penetration Testing.}
  33rd USENIX Security Symposium, 2024. arXiv:2308.06782. The arXiv
  preprint carries an earlier title; the version cited is the USENIX
  Security 2024 paper.
\item
  A. K. Zhang, N. Perry, R. Dulepet, J. Ji, C. Menders, J. W. Lin, E.
  Jones, G. Hussein, S. Liu, D. Jasper, P. Peetathawatchai, A. Glenn, V.
  Sivashankar, D. Zamoshchin, L. Glikbarg, D. Askaryar, M. Yang, T.
  Zhang, R. Alluri, N. Tran, R. Sangpisit, P. Yiorkadjis, K. Osele, G.
  Raghupathi, D. Boneh, D. E. Ho and P. Liang. \emph{Cybench: A
  Framework for Evaluating Cybersecurity Capabilities and Risks of
  Language Models.} ICLR, 2025. arXiv:2408.08926.
\item
  M. Shao, S. Jancheska, M. Udeshi, B. Dolan-Gavitt, H. Xi, K. Milner,
  B. Chen, M. Yin, S. Garg, P. Krishnamurthy, F. Khorrami, R. Karri and
  M. Shafique. \emph{NYU CTF Bench: A Scalable Open-Source Benchmark
  Dataset for Evaluating LLMs in Offensive Security.} NeurIPS Datasets
  and Benchmarks Track, 2024. arXiv:2406.05590.
\item
  P. K. Manadhata and J. M. Wing. \emph{An Attack Surface Metric.} IEEE
  Transactions on Software Engineering 37(3):371--386, 2011.
  doi:10.1109/TSE.2010.60.
\item
  N. Munaiah and A. Meneely. \emph{Attack Surface Definitions: A
  Systematic Literature Review.} Information and Software Technology
  104:94--103, 2019. doi:10.1016/j.infsof.2018.07.008.
\item
  K. Thompson. \emph{Reflections on Trusting Trust.} Communications of
  the ACM 27(8):761--763, 1984. doi:10.1145/358198.358210.
\item
  S. Torres-Arias, H. Afzali, T. K. Kuppusamy, R. Curtmola and J.
  Cappos. \emph{in-toto: Providing farm-to-table guarantees for bits and
  bytes.} 28th USENIX Security Symposium, 2019.
\item
  L. Zheng, W.-L. Chiang, Y. Sheng, S. Zhuang, Z. Wu, Y. Zhuang, Z. Lin,
  Z. Li, D. Li, E. P. Xing, H. Zhang, J. E. Gonzalez and I. Stoica.
  \emph{Judging LLM-as-a-Judge with MT-Bench and Chatbot Arena.} NeurIPS
  Datasets and Benchmarks Track, 2023. arXiv:2306.05685.
\item
  P. Wang, L. Li, L. Chen, Z. Cai, D. Zhu, B. Lin, Y. Cao, Q. Liu, T.
  Liu and Z. Sui. \emph{Large Language Models are not Fair Evaluators.}
  ACL, 2024. arXiv:2305.17926.
\item
  A. Panickssery, S. R. Bowman and S. Feng. \emph{LLM Evaluators
  Recognize and Favor Their Own Generations.} NeurIPS, 2024.
  arXiv:2404.13076.
\item
  D. Li, B. Jiang, L. Huang, A. Beigi, C. Zhao, Z. Tan, A.
  Bhattacharjee, Y. Jiang, C. Chen, T. Wu, K. Shu, L. Cheng and H. Liu.
  \emph{From Generation to Judgment: Opportunities and Challenges of
  LLM-as-a-judge.} arXiv preprint, 2024. arXiv:2411.16594.
\item
  X. Wang, J. Wei, D. Schuurmans, Q. Le, E. Chi, S. Narang, A. Chowdhery
  and D. Zhou. \emph{Self-Consistency Improves Chain of Thought
  Reasoning in Language Models.} ICLR, 2023. arXiv:2203.11171.
\item
  P. Verga, S. Hofstatter, S. Althammer, Y. Su, A. Piktus, A.
  Arkhangorodsky, M. Xu, N. White and P. Lewis. \emph{Replacing Judges
  with Juries: Evaluating LLM Generations with a Panel of Diverse
  Models.} arXiv preprint, 2024. arXiv:2404.18796.
\item
  Y. Zhao, H. Liu, D. Yu, S. Y. Kung, H. Mi and D. Yu. \emph{One Token
  to Fool LLM-as-a-Judge.} arXiv preprint, 2025. arXiv:2507.08794.
\item
  J. Shi, Z. Yuan, Y. Liu, Y. Huang, P. Zhou, L. Sun and N. Z. Gong.
  \emph{Optimization-based Prompt Injection Attack to LLM-as-a-Judge.}
  ACM CCS, 2024. arXiv:2403.17710.
\item
  K. Greshake, S. Abdelnabi, S. Mishra, C. Endres, T. Holz and M. Fritz.
  \emph{Not what you\textquotesingle ve signed up for: Compromising
  Real-World LLM-Integrated Applications with Indirect Prompt
  Injection.} ACM AISec, 2023. arXiv:2302.12173.
\item
  F. Perez and I. Ribeiro. \emph{Ignore Previous Prompt: Attack
  Techniques For Language Models.} NeurIPS ML Safety Workshop, 2022.
  arXiv:2211.09527.
\item
  Y. Liu, Y. Jia, R. Geng, J. Jia and N. Z. Gong. \emph{Formalizing and
  Benchmarking Prompt Injection Attacks and Defenses.} 33rd USENIX
  Security Symposium, 2024. arXiv:2310.12815.
\item
  E. Debenedetti, J. Zhang, M. Balunović, L. Beurer-Kellner, M. Fischer
  and F. Tramèr. \emph{AgentDojo: A Dynamic Environment to Evaluate
  Prompt Injection Attacks and Defenses for LLM Agents.} NeurIPS
  Datasets and Benchmarks Track, 2024. arXiv:2406.13352.
\item
  Q. Zhan, Z. Liang, Z. Ying and D. Kang. \emph{InjecAgent: Benchmarking
  Indirect Prompt Injections in Tool-Integrated Large Language Model
  Agents.} Findings of ACL, 2024. arXiv:2403.02691.
\item
  B. Biggio and F. Roli. \emph{Wild Patterns: Ten Years After the Rise
  of Adversarial Machine Learning.} Pattern Recognition 84:317--331,
  2018. arXiv:1712.03141.
\item
  N. Carlini and D. Wagner. \emph{Adversarial Examples Are Not Easily
  Detected: Bypassing Ten Detection Methods.} ACM AISec, 2017.
  arXiv:1705.07263.
\item
  F. Tramèr, N. Carlini, W. Brendel and A. Madry. \emph{On Adaptive
  Attacks to Adversarial Example Defenses.} NeurIPS, 2020.
  arXiv:2002.08347.
\item
  F. Pierazzi, F. Pendlebury, J. Cortellazzi and L. Cavallaro.
  \emph{Intriguing Properties of Adversarial ML Attacks in the Problem
  Space.} IEEE Symposium on Security and Privacy, 2020.
  arXiv:1911.02142.
\item
  F. Pendlebury, F. Pierazzi, R. Jordaney, J. Kinder and L. Cavallaro.
  \emph{TESSERACT: Eliminating Experimental Bias in Malware
  Classification across Space and Time.} 28th USENIX Security Symposium,
  2019. arXiv:1807.07838.
\item
  D. Arp, E. Quiring, F. Pendlebury, A. Warnecke, F. Pierazzi, C.
  Wressnegger, L. Cavallaro and K. Rieck. \emph{Dos and
  Don\textquotesingle ts of Machine Learning in Computer Security.} 31st
  USENIX Security Symposium, 2022. arXiv:2010.09470.
\item
  G. Klees, A. Ruef, B. Cooper, S. Wei and M. Hicks. \emph{Evaluating
  Fuzz Testing.} ACM CCS, 2018. arXiv:1808.09700.
\item
  A. Hazimeh, A. Herrera and M. Payer. \emph{Magma: A Ground-Truth
  Fuzzing Benchmark.} ACM SIGMETRICS, 2021. arXiv:2009.01120.
\item
  R. Croft, M. A. Babar and M. M. Kholoosi. \emph{Data Quality for
  Software Vulnerability Datasets.} ICSE, 2023. arXiv:2301.05456.
\item
  S. Chakraborty, R. Krishna, Y. Ding and B. Ray. \emph{Deep Learning
  based Vulnerability Detection: Are We There Yet?.} IEEE Transactions
  on Software Engineering, 2021. arXiv:2009.07235.
\item
  B. Steenhoek, M. M. Rahman, R. Jiles and W. Le. \emph{An Empirical
  Study of Deep Learning Models for Vulnerability Detection.} ICSE,
  2023. arXiv:2212.08109.
\item
  Y. Ding, Y. Fu, O. Ibrahim, C. Sitawarin, X. Chen, B. Alomair, D.
  Wagner, B. Ray and Y. Chen. \emph{Vulnerability Detection with Code
  Language Models: How Far Are We?.} ICSE, 2025. arXiv:2403.18624.
\item
  C. Rossow, C. J. Dietrich, C. Grier, C. Kreibich, V. Paxson, N.
  Pohlmann, H. Bos and M. van Steen. \emph{Prudent Practices for
  Designing Malware Experiments: Status Quo and Outlook.} IEEE Symposium
  on Security and Privacy, 2012. doi:10.1109/SP.2012.14.
\item
  T. Avgerinos, S. K. Cha, B. L. T. Hao and D. Brumley. \emph{AEG:
  Automatic Exploit Generation.} NDSS, 2011.
\item
  Y. Shoshitaishvili, R. Wang, C. Salls, N. Stephens, M. Polino, A.
  Dutcher, J. Grosen, S. Feng, C. Hauser, C. Kruegel and G. Vigna.
  \emph{SoK: (State of) The Art of War: Offensive Techniques in Binary
  Analysis.} IEEE Symposium on Security and Privacy, 2016.
  doi:10.1109/SP.2016.17.
\item
  R. Parasuraman and V. Riley. \emph{Humans and Automation: Use, Misuse,
  Disuse, Abuse.} Human Factors 39(2):230--253, 1997.
  doi:10.1518/001872097778543886.
\item
  G. Bansal, T. Wu, J. Zhou, R. Fok, B. Nushi, E. Kamar, M. T. Ribeiro
  and D. S. Weld. \emph{Does the Whole Exceed its Parts? The Effect of
  AI Explanations on Complementary Team Performance.} ACM CHI, 2021.
  arXiv:2006.14779.
\item
  Z. Buçinca, M. B. Malaya and K. Z. Gajos. \emph{To Trust or to Think:
  Cognitive Forcing Functions Can Reduce Overreliance on AI in
  AI-assisted Decision-making.} Proc. ACM Human-Computer Interaction
  5(CSCW1), 2021. arXiv:2102.09692.
\item
  M. Schemmer, N. Kuehl, C. Benz, A. Bartos and G. Satzger.
  \emph{Appropriate Reliance on AI Advice: Conceptualization and the
  Effect of Explanations.} ACM IUI, 2023. arXiv:2302.02187.
\item
  H. Vasconcelos, M. Jörke, M. Grunde-McLaughlin, T. Gerstenberg, M. S.
  Bernstein and R. Krishna. \emph{Explanations Can Reduce Overreliance
  on AI Systems During Decision-Making.} Proc. ACM Human-Computer
  Interaction 7(CSCW1), 2023. arXiv:2212.06823.
\item
  K. Zhou, J. Hwang, X. Ren and M. Sap. \emph{Relying on the Unreliable:
  The Impact of Language Models\textquotesingle{} Reluctance to Express
  Uncertainty.} ACL, 2024. arXiv:2401.06730.
\item
  J. L. Fleiss. \emph{Measuring Nominal Scale Agreement Among Many
  Raters.} Psychological Bulletin 76(5):378--382, 1971.
  doi:10.1037/h0031619.
\item
  J. R. Landis and G. G. Koch. \emph{The Measurement of Observer
  Agreement for Categorical Data.} Biometrics 33(1):159--174, 1977.
  doi:10.2307/2529310.
\item
  K. Krippendorff. \emph{Reliability in Content Analysis: Some Common
  Misconceptions and Recommendations.} Human Communication Research
  30(3):411--433, 2004. doi:10.1111/j.1468-2958.2004.tb00738.x.
\item
  D. Manheim and S. Garrabrant. \emph{Categorizing Variants of
  Goodhart\textquotesingle s Law.} arXiv preprint, 2018.
  arXiv:1803.04585.
\item
  D. Amodei, C. Olah, J. Steinhardt, P. Christiano, J. Schulman and D.
  Mané. \emph{Concrete Problems in AI Safety.} arXiv preprint, 2016.
  arXiv:1606.06565.
\item
  J. Skalse, N. H. R. Howe, D. Krasheninnikov and D. Krueger.
  \emph{Defining and Characterizing Reward Hacking.} NeurIPS, 2022.
  arXiv:2209.13085.
\item
  A. Pan, K. Bhatia and J. Steinhardt. \emph{The Effects of Reward
  Misspecification: Mapping and Mitigating Misaligned Models.} ICLR,
  2022. arXiv:2201.03544.
\item
  J. Jacobs, S. Romanosky, B. Edwards, M. Roytman and I. Adjerid.
  \emph{Exploit Prediction Scoring System (EPSS).} Digital Threats:
  Research and Practice 2(3), 2021. arXiv:1908.04856.
\item
  A. D. Householder, J. Chrabaszcz, T. Novelly, D. Warren and J. M.
  Spring. \emph{Historical Analysis of Exploit Availability Timelines.}
  USENIX CSET Workshop, 2020.
\item
  ProjectDiscovery. \emph{nuclei-templates: Community-curated detection
  templates for the Nuclei scanner.} GitHub repository,
  https://github.com/projectdiscovery/nuclei-templates (MIT licence;
  snapshot of 11 August 2026), 2026.
\item
  ProjectDiscovery. \emph{interactsh: An out-of-band interaction
  gathering server and client library.} GitHub repository,
  https://github.com/projectdiscovery/interactsh, 2026.
\end{enumerate}

\section{Supplementary material}\label{supplementary-material}

Supplementary Figures S1--S3 and Tables S1--S9 follow. Null results are
reported alongside the smallest attainable \(p\) of their design where
one exists.

\includegraphics[width=6.1in,height=4.86232in]{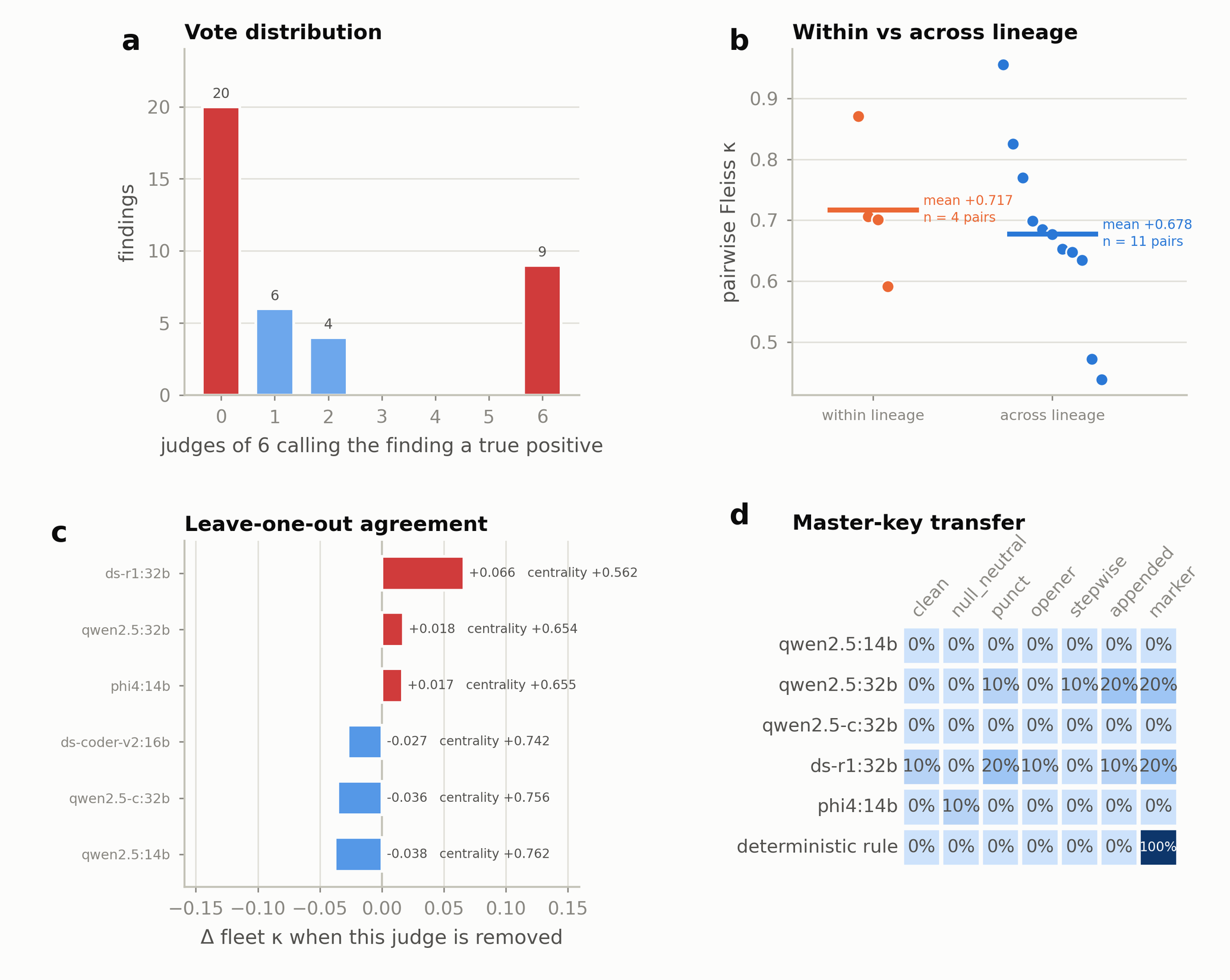}

\textbf{Figure S1, related to Figure 3. Agreement between model judges.}
\textbf{(a)} Distribution of the fleet vote over 39 findings and six
judges: the \(x\)-axis is the number of judges calling a finding a true
positive and the bar height the number of findings. Agreement is
substantial (Fleiss \(\kappa = + 0.683\), Krippendorff
\(\alpha = + 0.685\)) and bimodal: 74~\% of findings lie at the extremes
of the vote (red) and the middle is nearly empty. This is why majority
voting suppresses individual wobble in Figure~3d, since a member must
both wobble and be pivotal. \textbf{(b)} Pairwise \(\kappa\) for all 15
judge pairs, split by whether the pair shares a training lineage; dots
are pairs and horizontal lines group means. The difference, \(+ 0.040\)
(exact permutation \(p = 0.30\), 60 relabellings), is not significant.
\textbf{(c)} Change in fleet \(\kappa\) when each judge is removed,
annotated with that judge\textquotesingle s mean pairwise agreement with
the others; red bars mark judges whose removal increases agreement.
Removing one judge changes fleet \(\kappa\) by \(+ 0.066\), more than
the lineage difference in (b). Self-consistency does not predict
agreement (\(\rho = + 0.154\), \(p = 0.77\)). \textbf{(d)} Master-key
grid: five judges and the deterministic rule (rows) under seven
conditions (columns); cells give the share of ten cases confirmed as
true positives. Transfer to judges is weak (the most affected judge
flips on 20~\% and three of five never flip), whereas the deterministic
marker rule flips on 100~\%. This rule is the SQL-injection confirmer,
which is dormant on this authorisation evidence; Figure~4 therefore
attacks a mechanism that fires.

\includegraphics[width=6.1in,height=3.80145in]{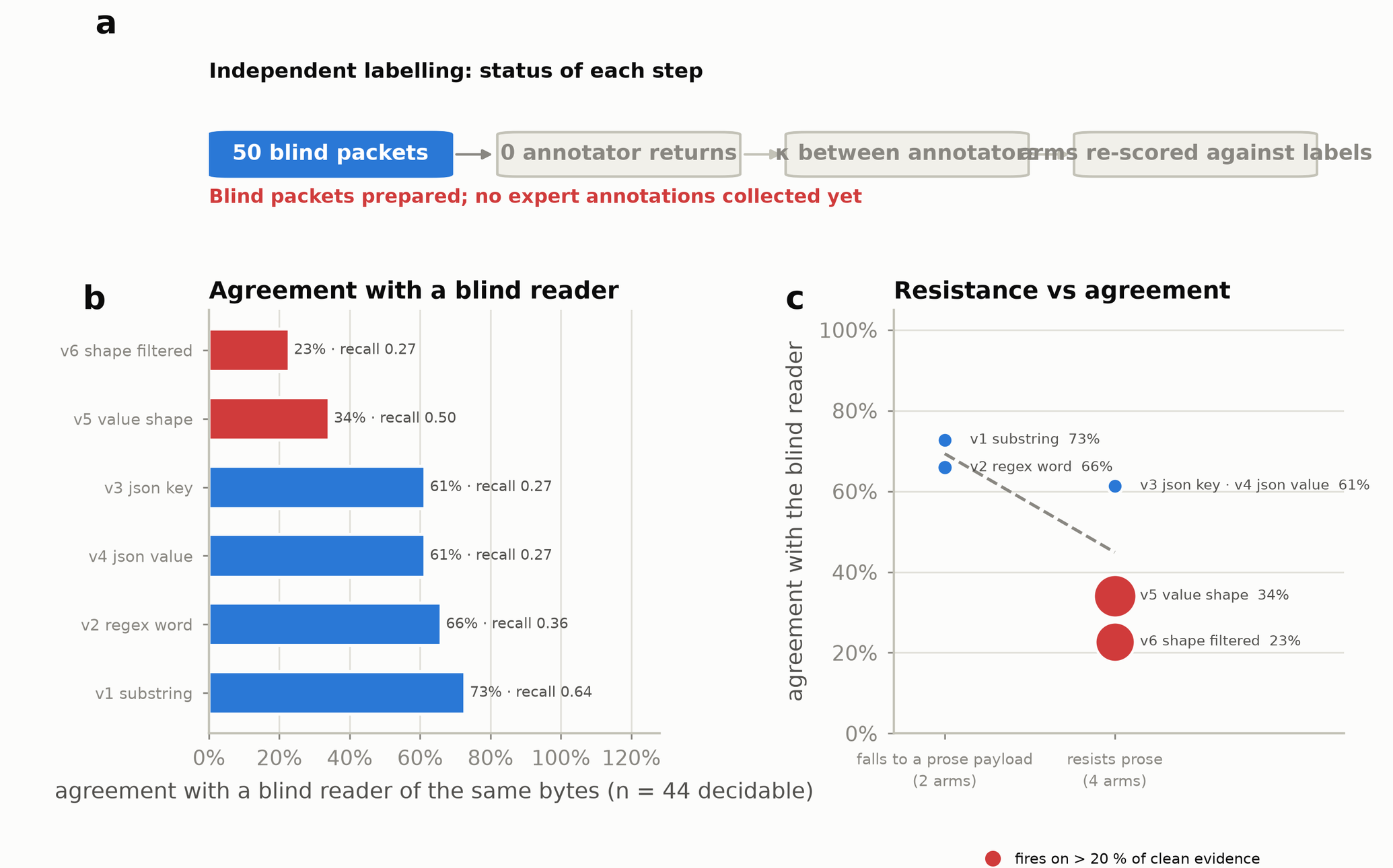}

\textbf{Figure S2, related to Figure 4 and the Limitations. Independent
labels.} All accuracies in the main figures are measured against labels
produced by the pipeline itself. \textbf{(a)} The independent-labelling
procedure in four steps: blinded packets drawn, annotations collected,
inter-annotator \(\kappa\) computed, arms re-scored. Only the first step
is complete: fifty findings were drawn from the pool of 200 with a fixed
seed, with five conclusion-bearing fields withheld, and the collection
workbook validates each return before analysis. Two expert annotators
are required for the remaining steps. \textbf{(b)} A machine-annotation
pilot, which is not ground truth, since an LLM cannot serve as the
reference in a study of the reliability of machine verdicts. It measures
agreement between each implementation and a blinded reader of the same
bytes over the 44 decidable packets. v1\_substring, which has 100~\%
sensitivity against the pipeline\textquotesingle s own labels by
construction, agrees with the reader on 73~\%; the two shape detectors
agree on 34~\% and 23~\% (red, below chance-level usefulness).
\textbf{(c)} The six implementations by resistance to a one-sentence
prose payload (\(x\)) and agreement with the blinded reader (\(y\));
marker area is proportional to the clean false-positive rate. Resistance
to forgery and agreement with the reader are inversely related. Six
packets are undecidable because redaction removes the deciding
information; in three cross-object reads the attack and baseline
requests become identical once the token is masked. This packet-builder
defect was detected by the pilot before expert annotation.

\includegraphics[width=6.1in,height=5.83478in]{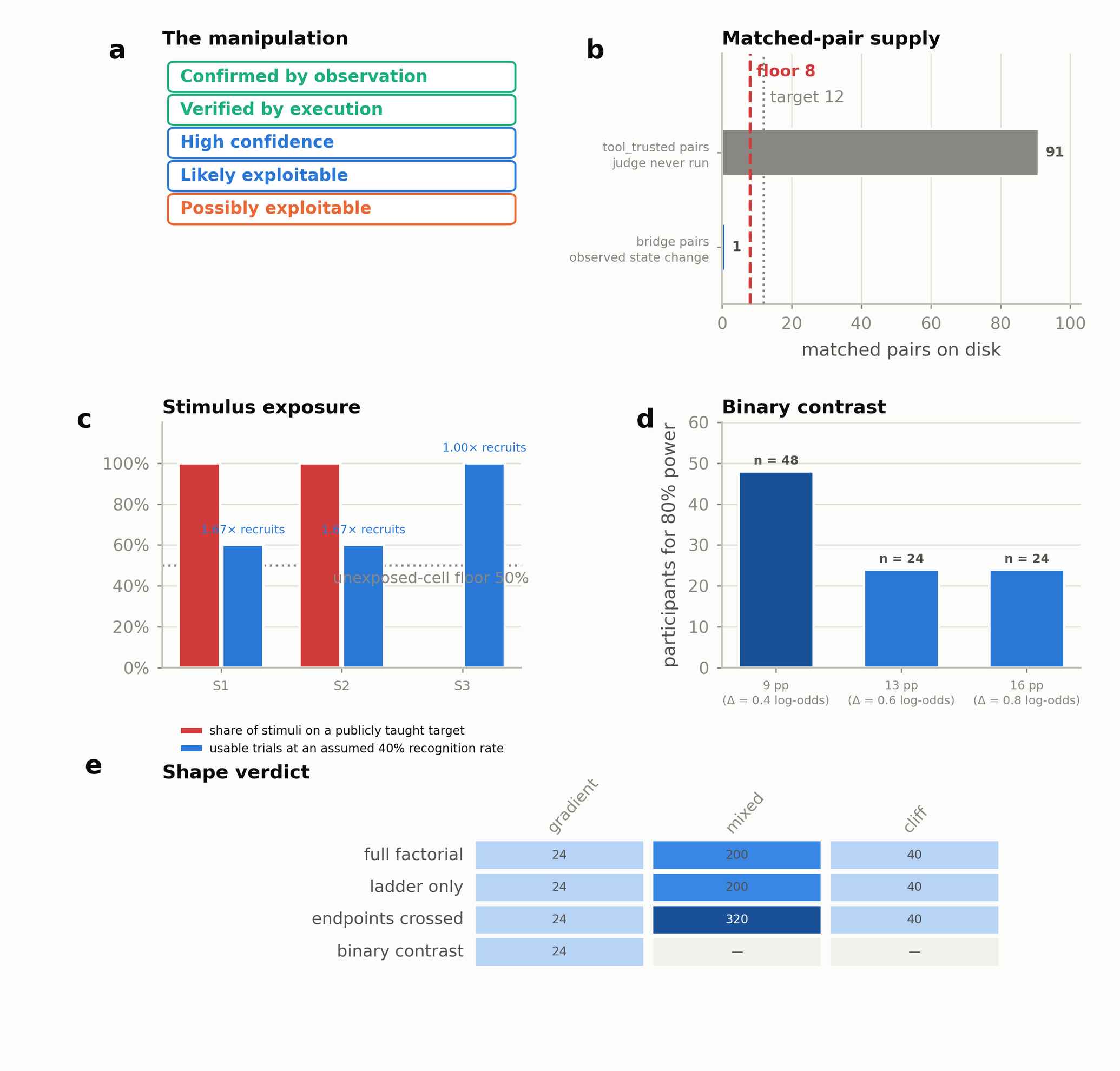}

\textbf{Figure S3, related to the Limitations. Design of the planned
operator study.} \textbf{(a)} The manipulation: a five-level scale of
asserted warrant, from a hedge (\emph{possibly exploitable}, orange)
through probability statements (blue) to assertions of an observed state
change (green). The quantities of interest are the acceptance gap
between deterministic and model-judged labels,
\(\Delta_{det} = P\left( \text{accept} \mid \text{det} \right) - P\left( \text{accept} \mid \text{model} \right)\),
the corresponding accuracy gap \(\Delta_{just}\), and their difference
\(\Delta_{excess} = \Delta_{det} - \Delta_{just}\), i.e. acceptance
beyond what accuracy justifies. \textbf{(b)} Supply of matched pairs, in
which the same finding was both model-judged and deterministically
confirmed: the run records contain \textbf{one}, against a
pre-registered minimum of eight (red dashed) and a target of twelve
(dotted). A further 91 pairs (grey), in which a
detector\textquotesingle s finding was kept without running the judge,
do not involve an observed state change and are not eligible.
\textbf{(c)} Stimulus exposure by study. Red, share of candidate stimuli
on a publicly taught target; blue, share of trials retained at an
assumed 40~\% recognition rate, annotated with the resulting recruitment
multiplier. All candidates for S1 and S2 are exposed, so provenance
cannot be separated from recognition. \textbf{(d)} Participants required
for 80~\% power on the binary contrast, by effect size, from 2,000
simulations per cell with participant and item random effects.
\textbf{(e)} Participants required for 80~\% power on the \emph{shape}
of the effect (gradient, cliff or mixed), by design (rows) and true
regime (columns); a dash marks a cell the design cannot identify at any
sample size. With two levels, the cliff indicator is an affine function
of the manipulation, so the binary contrast cannot distinguish a
gradient from a step. \textbf{No human-subject data have been collected,
and no effect size is assumed.} Design targets are listed in Table~S9.

\section{Supplementary tables}\label{supplementary-tables}

\textbf{Table S1. Failure geometry, arm by arm.} All arms measured on
the same 147-case clean-negative pool over 13 attacker budgets. AUFC is
the area under the forgeability curve; $\alpha^\ast$ is the budget at which the arm
breaks.

\includegraphics[width=6.1in,height=2.11482in]{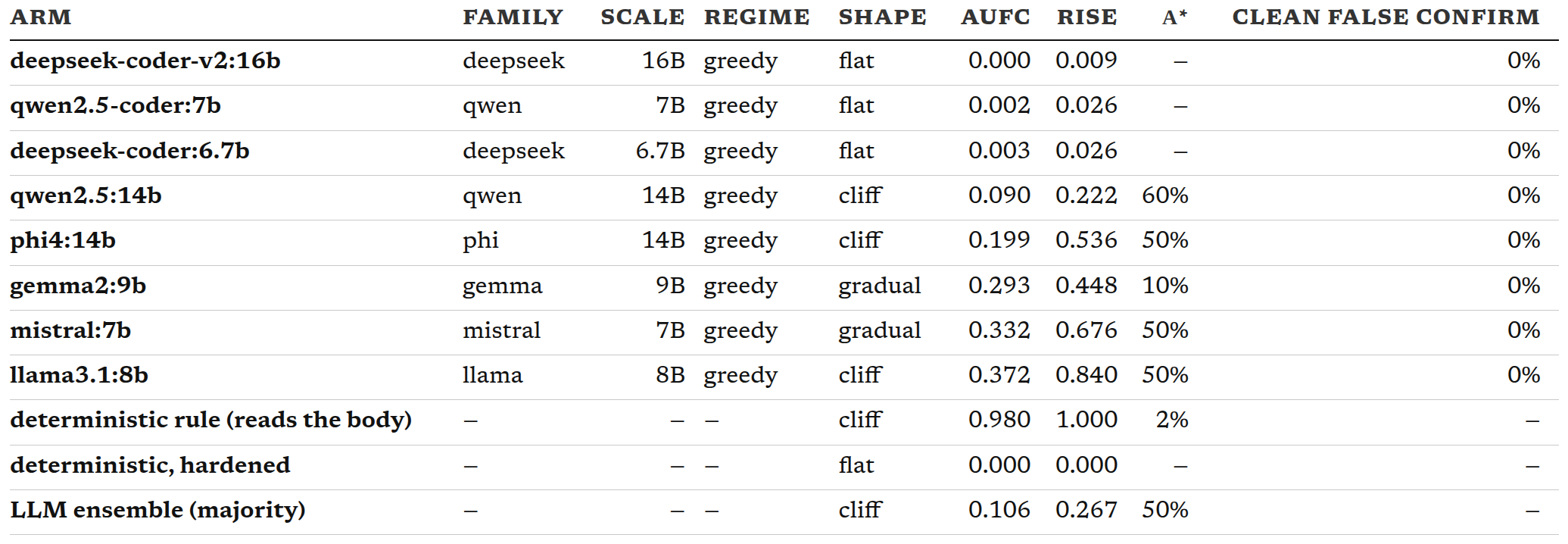}

Note. $\rho$(parameters, AUFC) = -0.193, exact p = 0.654. The within-scale
control is the 5 arms in the 6.7--9B band, drawn from 5 families.
Separation margin from the deterministic rule: 0.607. Four arms
evaluated only on a 10-case pilot pool are excluded, because areas over
10 and 147 cases are not comparable.

\textbf{Table S2. Minimum attacker budget b*, by adaptive search.}
Generate-then-minimise search over seven payload families, not a fixed
ladder. Every b* is an upper bound: a cleverer attacker pays no more.
Adversary model: append.

\includegraphics[width=6.1in,height=4.21787in]{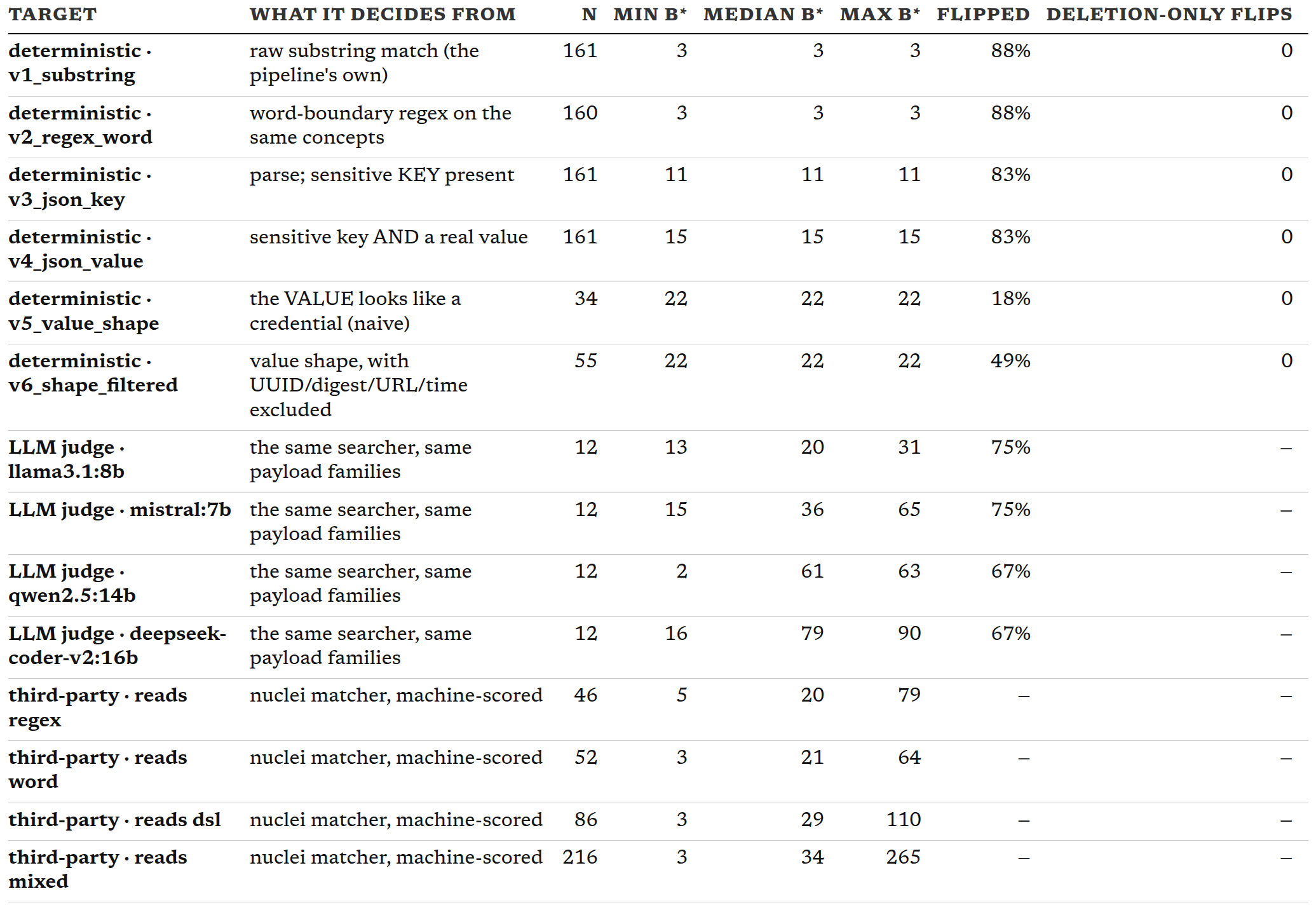}

Note. Each deterministic arm is searched only on its own clean
negatives. The deletion-only control (truncation without added content)
flips no case. The prompt-injection payload family wins no case against
any deterministic arm.

\textbf{Table S3. Six independent implementations of one check under six
attack conditions.} confirm\_unauth fires on 39 of 200 evidence
instances and declines on the remaining 161, which form the attack set.
Each cell is the share of attackable cases flipped; the clean column is
the unattacked control.

\includegraphics[width=6.1in,height=2.03235in]{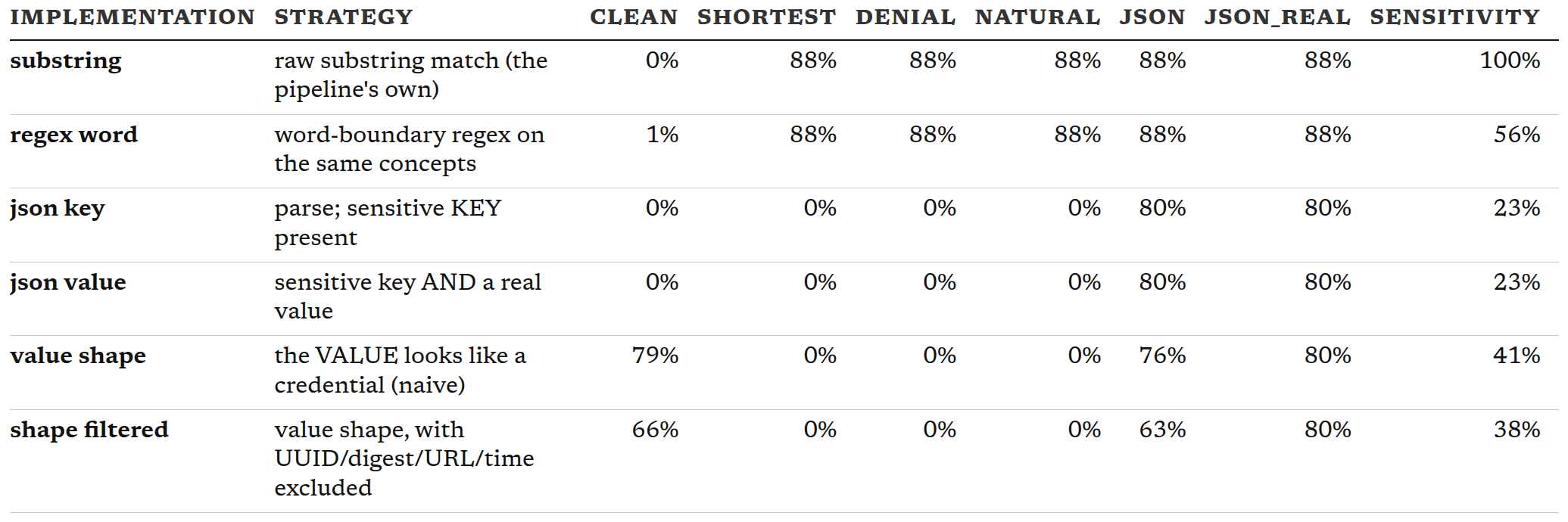}

Note. Worst condition: json\_real. The two shape detectors resist prose
payloads but already fire on most unmodified evidence (clean column).

\textbf{Table S4. Activation of the status-inversion branch, six
implementations.} 200 instances. The shipped rule returns inconclusive
on all of them through both branches; the table therefore reports
activation, i.e. how many attacker-authored fields are needed to satisfy
the conjunction.

\includegraphics[width=6.1in,height=3.5903in]{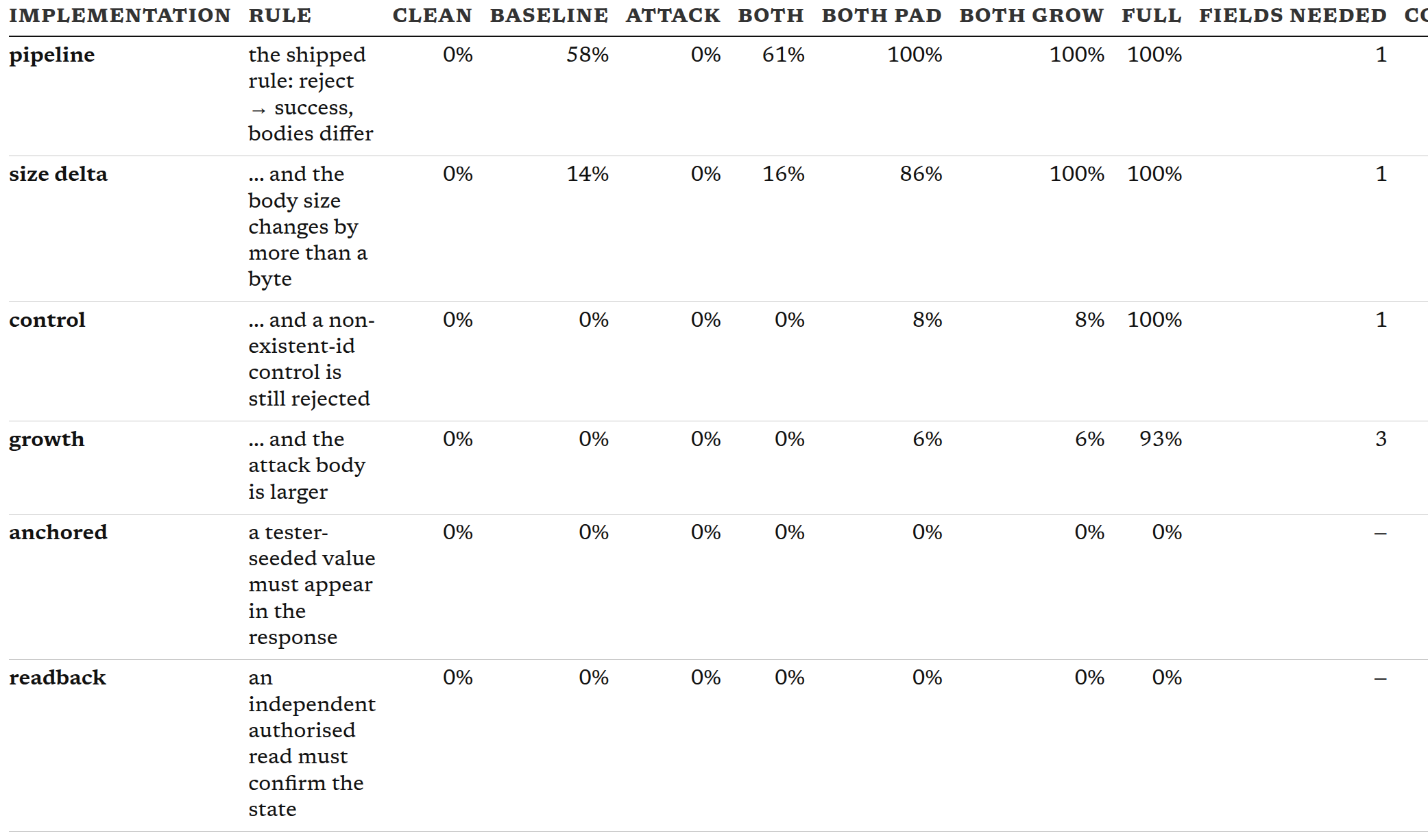}

Note. A conjunction costs the attacker only the conjuncts that are not
already true. The anchored and read-back implementations cannot be
activated.

\textbf{Table S5. Confirmation policies, binary and ternary.} Measured
on 5 constructed positives, 39 observed positives, 27 clean negatives
and 1,932 attacked negatives. The two sensitivity columns use different
denominators.

\includegraphics[width=6.1in,height=2.28128in]{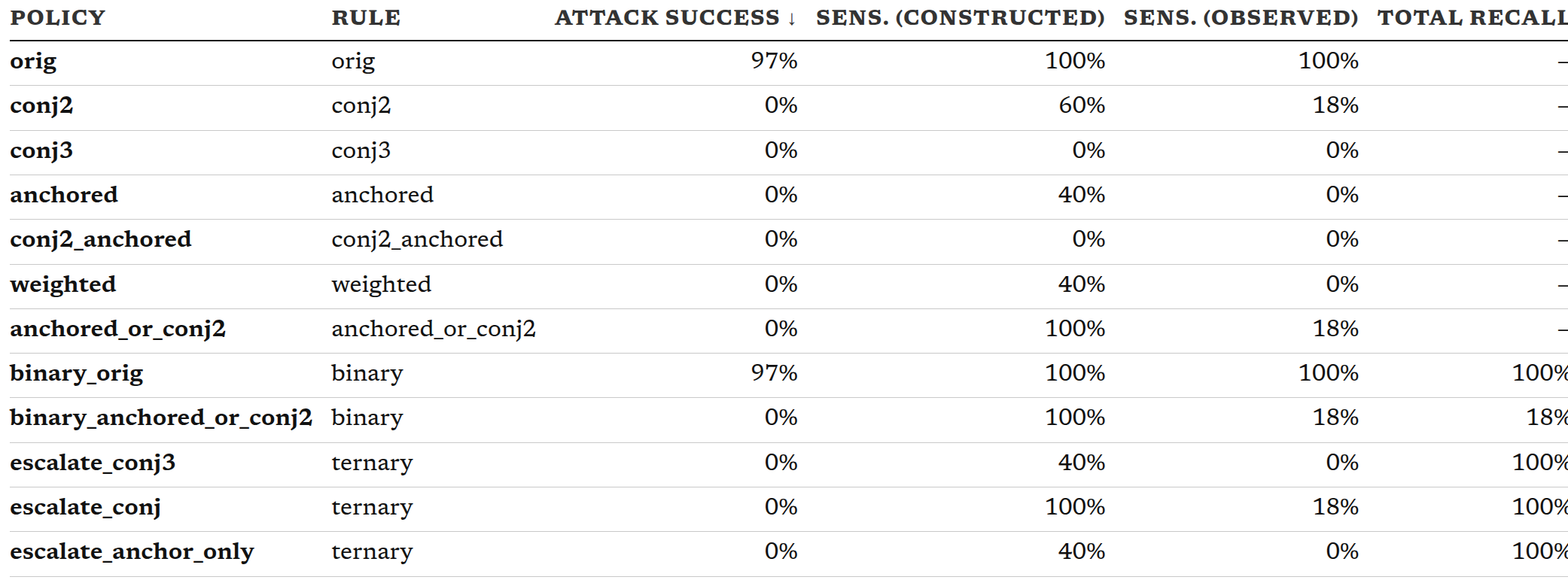}

Note. Frontier membership uses both sensitivity axes; scoring confirmed
sensitivity alone would treat an escalated positive as a missed one.
Escalation on clean evidence is 0 \% for every ternary policy.

\textbf{Table S6. Ensemble statistics by fleet size and roster.} $\kappa$
computed over 39 findings, sensitivity over 31 observed positives.
Instability is the share of (roster, case) pairs whose verdict differs
from the full fleet\textquotesingle s on identical evidence.

\includegraphics[width=6.1in,height=0.8929in]{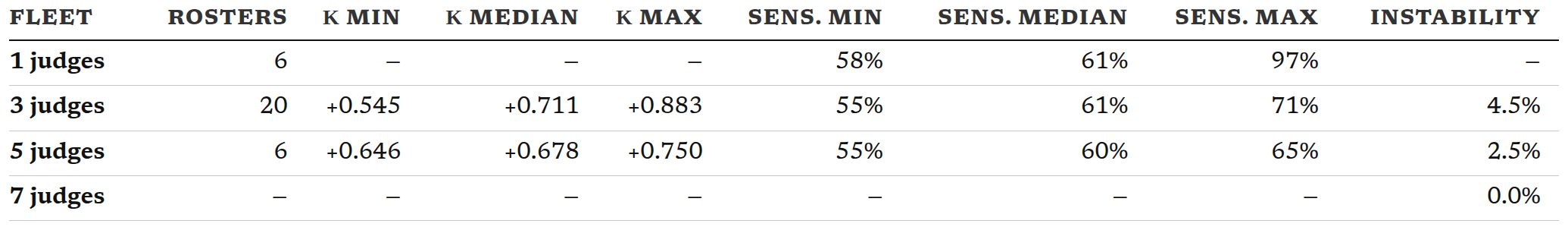}

Note. The full fleets give $\kappa$ = +0.683 (6 judges) and sensitivity 54.8\%
(6 judges); a different roster of the same size changes both. The 0 \%
instability at seven judges reflects a single possible roster. The same
fleet queried twice is 99.7\% self-consistent over 2,380 inferences.

\textbf{Table S7. The reachability gate applied to a third-party
template corpus.} Each mechanism is scored automatically from the
template\textquotesingle s matcher specification (nuclei-templates). The
two blocks differ only in the adversary model.

\includegraphics[width=6.1in,height=4.07648in]{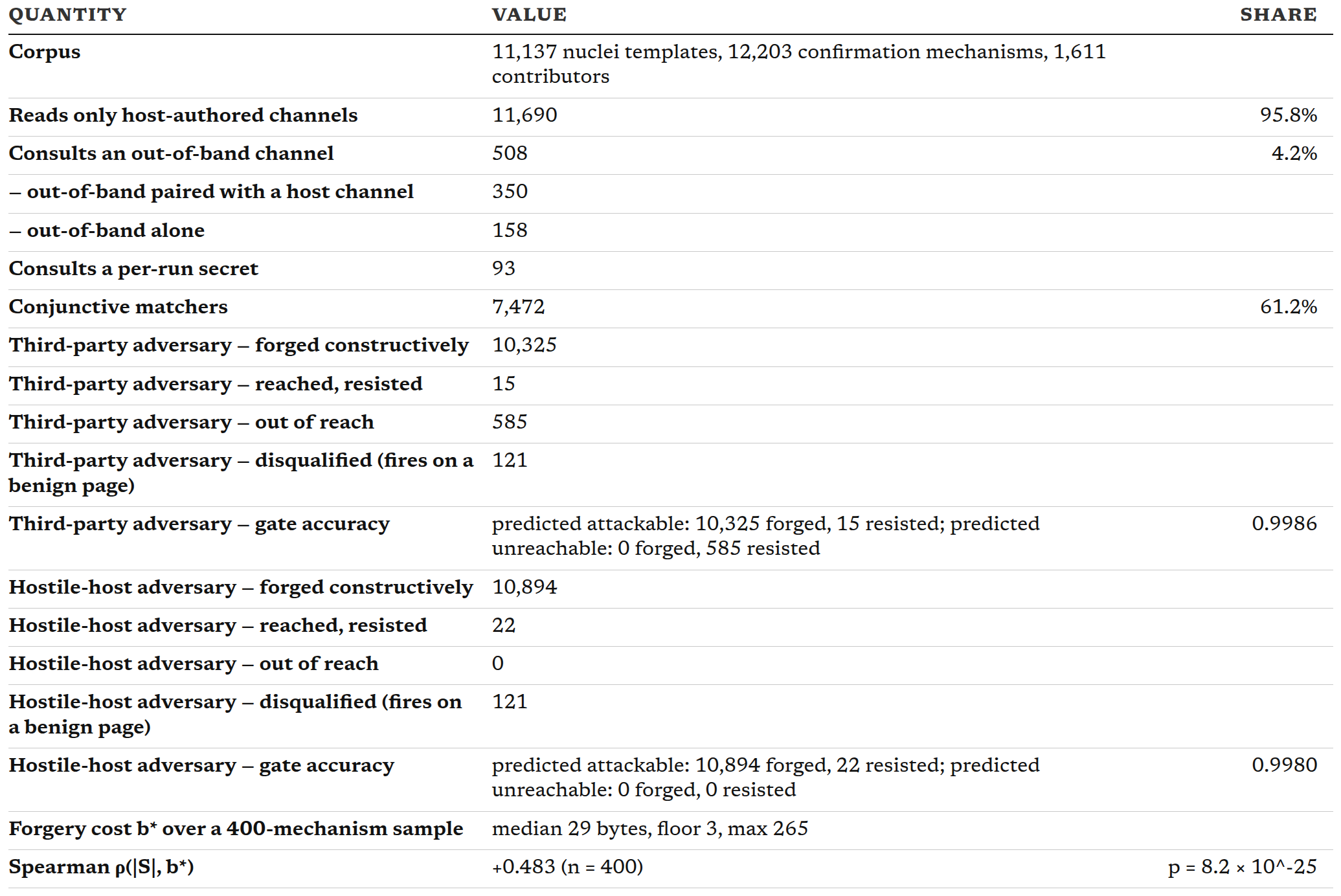}

Note. Under the third-party adversary, the 585 unreachable mechanisms
resist by definition, since the attacker cannot write the channel they
read. The scanned host, by contrast, receives the out-of-band callback
URL in the payload and the per-run value in the request, so under the
hostile-host adversary no mechanism is out of reach.

\textbf{Table S8. Per-judge statistics of the agreement fleet.} Fleet
Fleiss $\kappa$ = +0.683 and Krippendorff $\alpha$ = +0.685 over 39 findings and 6
judges. 74\% of findings sit at an extreme of the vote. Centrality is a
judge\textquotesingle s mean pairwise $\kappa$ against the rest of the fleet;
$\Delta\kappa$ is what removing it does to the fleet.

\includegraphics[width=6.1in,height=1.68387in]{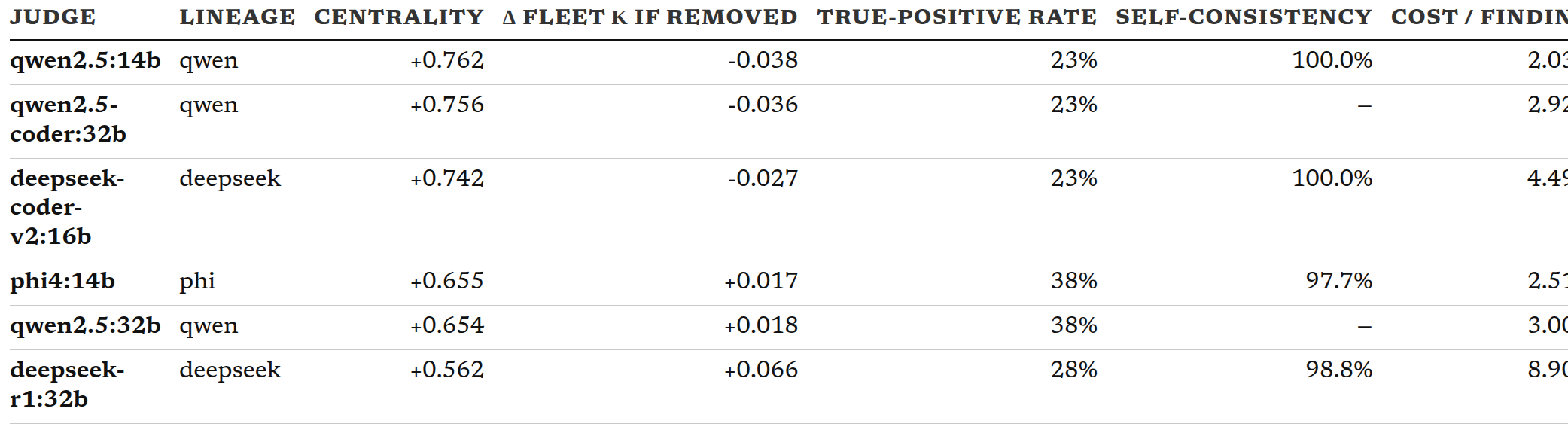}

Note. Within- minus across-lineage $\kappa$ is +0.040 (exact permutation p =
0.30); self-consistency does not predict agreement ($\rho$ = +0.154, p =
0.77). Self-consistency is from 2,380 repeated inferences; a dash marks
a judge not in the repeatability study.

\textbf{Table S9. Planned operator study: supply, stimuli and power.} No
human-subject data have been collected. Values are design targets and
simulated sample-size requirements, not results.

\includegraphics[width=6.1in,height=4.65309in]{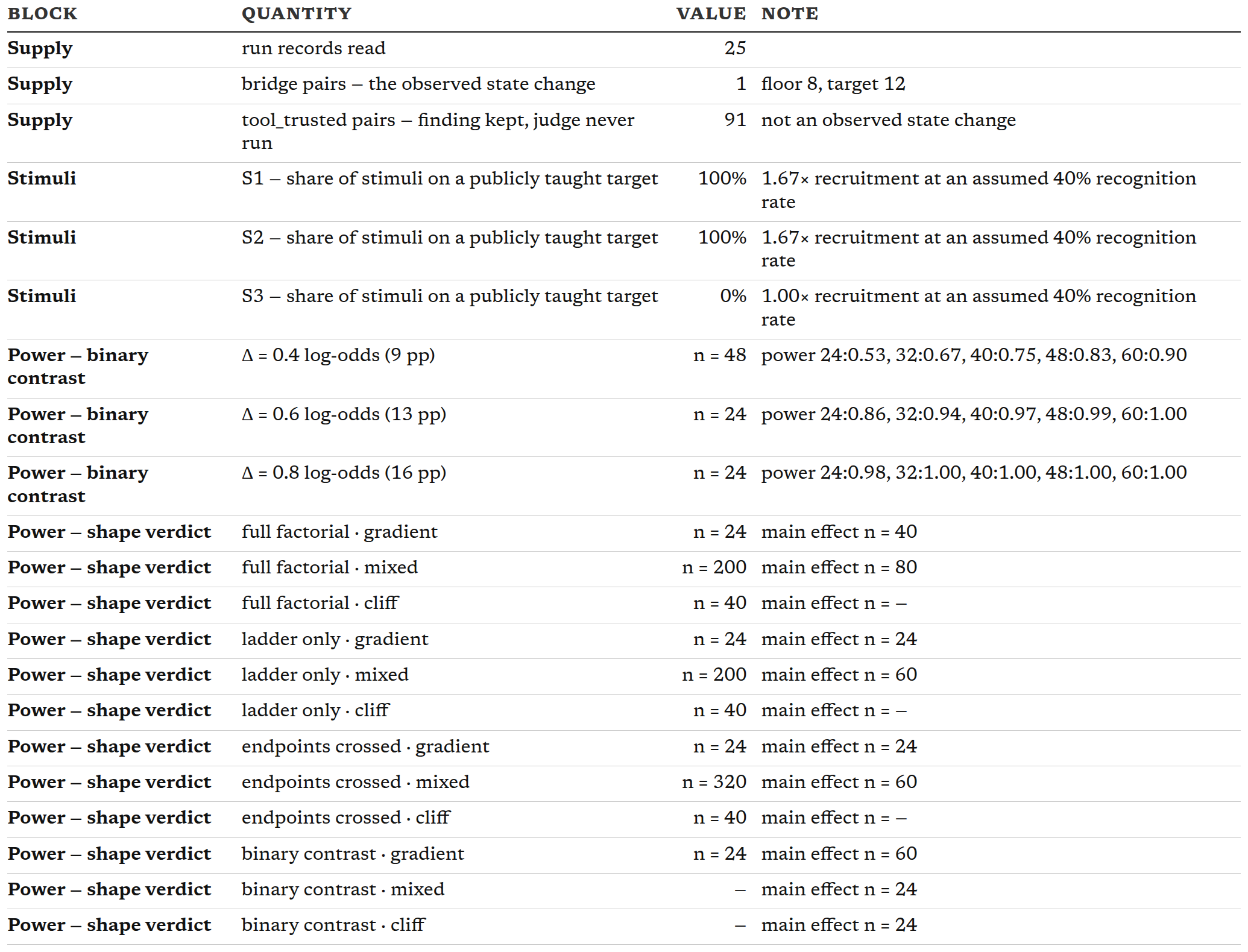}

Note. The study is blocked by supply: the pre-registered matched-pair
floor is not met, and every candidate stimulus lies on a publicly taught
target, so provenance cannot be separated from recognition without a
non-public target.

\end{document}